# STeP-CiM: <u>S</u>train-enabled <u>T</u>ernary <u>P</u>recision <u>C</u>omputation-<u>i</u>n-<u>M</u>emory based on Non-Volatile 2D Piezoelectric Transistors

**Niharika Thakuria[1*], Reena Elangovan[1], Sandeep K Thirumala[1,2], Anand Raghunathan[1], Sumeet K. Gupta[1]**

[1]School of Electrical and Computer Engineering, Purdue University, West Lafayette, IN, USA

[2] Currently at Intel Corporation, Santa Clara, CA, USA

**\* Correspondence:**
Niharika Thakuria
nthakuri@purdue.edu

**Abstract**

We propose 2D Piezoelectric FET (PeFET) based compute-enabled non-volatile memory for ternary deep neural networks (DNNs). PeFETs hinge on ferroelectricity for bit storage and piezoelectricity for bit sensing, exhibiting inherently amenable features for computation-in-memory of dot products of weights and inputs in the signed ternary regime. PeFETs consist of a material with ferroelectric and piezoelectric properties coupled with Transition Metal Dichalcogenide channel. We utilize (a) ferroelectricity to store binary bits (0/1) in the form of polarization (-$P$/+$P$) and (b) polarization dependent piezoelectricity to read the stored state by means of strain-induced bandgap change in Transition Metal Dichalcogenide channel. The unique read mechanism of PeFETs enables us to expand the traditional association of +$P$ (-$P$) with low (high) resistance states to their dual high (low) resistance depending on read voltage. Specifically, we demonstrate that +$P$ (-$P$) stored in PeFETs can be dynamically configured in (a) a low (high) resistance state for positive read voltages and (b) their dual high (low) resistance states for negative read voltages, without afflicting a read disturb. Such a feature, which we name as Polarization Preserved Piezoelectric Effect Reversal with Dual Voltage Polarity (PiER), is unique to PeFETs and has not been shown in hitherto explored memories. We leverage PiER to propose a <u>S</u>train-enabled <u>T</u>ernary <u>P</u>recision <u>C</u>omputation-<u>i</u>n-<u>M</u>emory (STeP-CiM) cell with capabilities of computing the scalar product of the stored weight and input, both of which are represented with signed ternary precision. Further, using multi word-line assertion of STeP-CiM cells, we achieve massively parallel computation of dot products of signed ternary inputs and weights. Our array level analysis shows 91% lower delay and improvements of 15% and 91% in energy for in-memory multiply-and-accumulate operations compared to near-memory design approaches based on 2D FET based SRAM and PeFET respectively. We also analyze the system-level implications of STeP-CiM by deploying it in a ternary DNN accelerator. STeP-CiM exhibits 6.11× - 8.91× average improvement in performance and 3.2× average improvement in energy over SRAM based near-memory design. We also compare STeP-CiM to near-memory design based on PeFETs showing 5.67× - 6.13× average performance improvement and 6.07× average energy savings.

## 1 Introduction

Deep Neural Networks (DNNs) have transformed the field of machine learning and are deployed in many real-world products and services (Lecun et al., 2015). However, enormous storage and compute demands limits their application in energy-constrained edge-devices (Venkataramani et al., 2016). Precision reduction in DNNs has emerged as a popular approach for energy-efficient realization of hardware accelerators for these applications (Courbariauxécole and Bengio, 2015; Mishra et al., 2017; Choi et al., 2018; Colangelo et al., 2018; Wang et al., 2018). State-of-the-art DNN hardware for inference employs 8-bit precision, and recent algorithmic efforts have shown the pathway for aggressive scaling up to binary precision (Choi et al., 2018; Colangelo et al., 2018) . However, accuracy suffers significantly at binary precision. Interestingly, ternary precision networks offer a near-optimal design point in the low precision regime with significant accuracy boost compared to binary DNNs (Li et al., 2016; Zhu et al., 2016) and large energy savings with mild accuracy loss compared to higher precision DNNs (Mishra et al., 2017; Wang et al., 2018). Due to these features, ternary precision networks have garnered interest for their hardware realizations (Jain et al., 2020; Thirumala et al., 2020). Ternary DNNs can be implemented using classical accelerator architectures (e.g., Tensor Processing Unit and Graphical Processing Unit) by employing specialized processing elements and on-chip scratchpads to improve energy efficiency, but are nevertheless limited by memory bottleneck. In this regard, computing-in-memory (CiM) brings a new opportunity that can greatly enhance efficiency of DNN accelerators by reducing power-hungry data transfer between memory and processors.

## 1.1 Related Works on Low Precision Computing-In-Memory (CiM) for DNNs

Several previous works have explored hardware realization of low-precision CiM for DNN workloads. For example, binary networks such as XNOR-RRAM (Sun et al., 2018) and XNOR-SRAM (Yin et al., 2020) feature large parallel vector-matrix multiplication capability, but suffer from low accuracies due to aggressive quantization of weights and inputs to binary values. At the other end of the spectrum, DNNs with 4-8 bits have attained high accuracies, albeit at the cost of considerably increased energy consumption and reduction in throughput (Liu et al., 2015; Chi et al., 2016). In this regard, ternary DNNs are attractive as they achieve a remarkably large upswing in accuracy compared to the binary networks while significantly reducing the energy consumption compared to higher precision networks (Mishra et al., 2017; Wang et al., 2018). In other words, ternary DNNs yield a near-optimal design point in the context of energy-accuracy trade-offs for energy-constrained applications, which has motivated several ternary CiM designs. (Yoo et al., 2019) proposed eDRAM based ternary CiM. However, the repetitive refresh operations add burden to the energy constrained edge devices. Emerging technologies such as resistive RAM (RRAM) (Chen et al., 2018; Liu et al., 2020; Doevenspeck et al., 2021), spin transfer/orbit torque magnetic RAM (STT/SOT-MRAM) (Doevenspeck et al., 2020; Bian et al., 2021) are also being actively explored for ternary precision networks due to their high density and low leakage power. However, their power-hungry current driven write (Si et al., 2021) lowers their favorability as a candidate for ternary CiM hardware targeted for energy constrained environments. The common aspect in all the above works is that they use signed ternary weights with binary inputs and do not attempt to exploit the accuracy benefits of pure signed ternary networks, i.e. with weights = {-1, 0, 1} and inputs ={-1, 0, 1}. Recent works have brought attention to hardware accelerator designs for pure signed ternary regime with static random access memory (SRAM) and non-volatile ferroelectric transistor based DNN architectures (Jain et al., 2020; Thirumala et al., 2020). These works report high parallelism, low energy and small accuracy loss, making a case for hardware architectures for signed ternary CiM. However, a downside of both designs is the requirement of hardware additions for achieving ternary CiM functionality. SRAM based ternary CiM implementations such as by (Jain et al., 2020) raise concerns for area efficiency and leakage energy. The use of non-volatile ferroelectric transistors in the ternary CiM design (Thirumala et al., 2020) remits area cost and leakage energy. However, existing ferroelectric based non-volatile memories suffer from other disadvantages that are discussed subsequently.

## 1.2 Background of Ferroelectric based Memories

Ferroelectric RAM or FERAM (Kim et al., 2007) is one of the earliest memories based on ferroelectric materials. It utilizes a ferroelectric capacitor along with an access transistor in a 1T-1C configuration. FERAMs feature high density, large endurance, high retention and electric field-driven write, which is more energy efficient compared to current-based write in other non-volatile memories (Si et al., 2021). However, it suffers from issues such as destructive read and low distinguishability between the memory states. Ferroelectric FETs (FEFETs), in which the ferroelectric material is integrated within the gate stack of a transistor (Yu et al., 2021), offer appealing attributes that mitigate the concerns of FERAMs. For instance, FEFETs feature separation of read-write paths, non-destructive read and high distinguishability, while retaining the benefits of electric-field driven write (Yu et al., 2021) and offering other advantages such as multi-level storage (Ni et al., 2018; Kazemi et al., 2020; Dutta et al., 2020; Liao et al., 2021). However, they are known to suffer from variability, endurance and retention concerns due to traps at the ferroelectric-dielectric interface and depolarization fields in the ferroelectric. Moreover, it is challenging to scale their write voltage. In order to achieve write voltage reduction, ferroelectric-metal-FETs (FEMFETs) were proposed by (Ni et al., 2018; Kazemi et al., 2020) which connect a ferroelectric capacitor with the gate of a transistor, allowing independent optimization of the cross-sectional area of two components. This is helpful in scaling the write voltage to logic-compatible levels. The ferroelectric capacitor can be formed directly on the gate stack or at the back-end of the line. Besides write-voltage reduction, FEMFETs mitigate the variability concerns of FEFETs due to the presence of metal between the ferroelectric and the dielectric of the transistor, which addresses the trap-related issues (Ni et al., 2018; Kazemi et al., 2020). However, this inter-layer metal (ILM) is floating and therefore, is susceptible to potential changes due to gate leakage, which leads to bit-sensing challenges (Thirumala and Gupta, 2018).

To address the issues of FERAM, FEFETs and FEMFETs, while still retaining the advantages of electric-field-driven write, we (Thakuria et al., 2020) had explored another flavor of a ferroelectric material-based memory called Piezoelectric FET (PeFET). PeFET utilizes both ferroelectric and piezoelectric properties of the ferroelectric material. PeFET consists of a ferroelectric capacitor coupled with a 2D Transition Metal Dichalcogenide (TMD) FET in a four-terminal structure with gate, drain, source and back contacts. The capacitor is designed with a material exhibiting strong ferroelectric and piezoelectric properties. PeFET utilizes polarization retention of the ferroelectric capacitor for bit storage. Its write operation involves applying suitable voltage across the ferroelectric capacitor to switch the polarization, similar to that of an FERAM. Therefore, PeFETs inherit the advantages of low power electric field driven switching, large endurance and high retention. Also, since the ferroelectric layer is controlled by metal layers on both ends, it does not suffer from severe trap-related issues observed in FEFETs. For read, PeFETs employ a unique mechanism based on dynamic bandgap change in the TMD FET induced by voltage-dependent strain of the ferroelectric/piezoelectric capacitor. This leads non-destructive read and separation of read-write paths (discussed later). Furthermore, there is no floating metal in PeFETs (unlike FEMFETs). This prevents issues related to gate leakage. One design challenge in PeFETs is limited distinguishability, which can be improved by choosing ferroelectric material exhibiting high piezoelectricity, e.g., PZT-5H (Malakooti and Sodano, 2013) and TMD material with high sensitivity of bandgap change to pressure, e.g. $MoS_2$ (Peña-Álvarez et al., 2015) and geometry optimization such as hammer and nail effect (Newns et al., 2012) to focus the strain on the TMD channel. These aspects are discussed in detail later. In summary, PeFETs address several important challenges observed in existing ferroelectric based memories while retaining the key advantage of electric-field driven write. In addition, as proposed in this work, they exhibit unique properties associated with polarization-induced strain that make them amenable for designing compute-enabled memories in the pure signed ternary regime.

### 1.3 Previous works on Piezoelectric based FETs

Initial proposals of piezoelectric based FETs were made in the context of steep-switching devices (Newns et al., 2012; Hueting et al., 2015; Das, 2016; Wang et al., 2018; Alidoosty-Shahraki et al., 2019). A material with high piezoelectric coefficient, such as Lead Magnesium Niobate-Lead Titanate (commonly known as PMN-PT) is utilized in such devices to modulate the resistance of a piezoresistive material (Newns et al., 2012) or bandgap of Si/TMD channel (Hueting et al., 2015; Das, 2016; Alidoosty-Shahraki et al., 2019). Our proposal of PeFET (Thakuria et al., 2020) extends the idea of piezoelectricity driven bandgap modulation of TMD beyond steep-switching devices to non-volatile memory (NVM) design. As already introduced, it stores bit information in a piezoelectric/ferroelectric material and leverages polarization-dependent piezoelectric response to modulate the bandgap of the TMD channel for sensing. As shown by (Thakuria et al., 2020) and discussed later, positive ferroelectric polarization (+$P$) leads to bandgap reduction in TMD and thus, low resistance state (LRS). On the other hand, negative polarization (-$P$) yields bandgap increase and high resistance state (HRS). The drain current of PeFET can be used to sense of the memory. Contrary to previous proposals of piezoelectric based FETs, PeFET NVM uses Lead Zirconate Titanate (PZT-5H) as piezoelectric (PE) to satisfy the following requirements: (i) sufficiently wide hysteresis of polarization-voltage response for non-volatile memory functionality (ferroelectric property) and (ii) large strain-voltage characteristics (piezoelectric property) for achieving effective bandgap modulation in TMD NVM. Various experiments have demonstrated monotonic bandgap reduction in TMD on the application of out-of-plane pressure (Nayak et al., 2014; Peña-Álvarez et al., 2015). For example, multilayer Molybdenum Disulfide ($MoS_2$) subjected to out-of-plane uniaxial stress has experimentally shown a bandgap reduction of ~80meV/GPa and achieves semiconductor-to-metal transistor at ~20GPa (Nayak et al., 2014). Monolayer $MoS_2$ achieves bandgap reduction of up to ~800meV/GPa (Peña-Álvarez et al., 2015). We use monolayer $MoS_2$ in this work due to its high bandgap coefficient.

### 1.4 Contributions of this work

In this paper, we identify that the unique read mechanism of PeFET can be extended beyond standard memory implementation proposed in (Thakuria et al., 2020). We build on this understanding to present PeFET enabled signed ternary CiM design. The key contributions of this paper are as follows:

1. We establish through simulations that LRS of +$P$ can be swapped to HRS while HRS of -$P$ to LRS by reversing the polarity of applied voltage across the piezoelectric during sensing. We name this feature as Polarization Preserved Piezoelectric Effect Reversal with Dual Voltage Polarity (PiER).

2. We explore PiER for ternary input encoding. We show that PiER motivates exploration of PeFET based non-volatile memory that naturally supports signed ternary CiM.

3. We propose a ternary compute-enabled non-volatile memory (STeP-CiM) using PeFET and PiER functionality that performs scalar multiplication of signed inputs and weights without extra transistors.

3. We show parallel in-memory dot-product computation with STeP-CiM based on current sensing, as opposed to voltage sensing in the previous ternary designs by (Jain et al., 2020; Thirumala et al., 2020). We discuss the implications of current sensing for signed ternary CiM, evaluate the energy and delay of STeP-CiM in comparison to near memory (NM) baselines based on PeFET (PeFET-NM) and SRAM (SRAM-NM).

4. We evaluate the system-level implications of STeP-CiM by implementing it in a DNN accelerator and quantify its energy and performance benefits over PeFET-NM and SRAM-NM baseline designs.

## 2 Device Structure, Materials, Methods of Modeling and Simulation

### 2.1 Device Structure and Operation of PeFET

PeFET is a four-terminal non-volatile device consisting of drain (*D*), gate (*G*), source (*S*) and back (*B*) contacts. We present the structure and schematic of a PeFET device in Figure 1(A, B). Its non-volatility is enabled by a ferroelectric material (PE) positioned between *G* and *B*, which also function as the write port of the device as illustrated in Figure 1(A). In addition to ferroelectricity, PE, which is PZT-5H in this work, exhibits good piezoelectric response (high piezoelectric coefficient value, $d_{33}$ = 650pm/V (Malakooti and Sodano, 2013) for successful sensing. On the other side of *G*, an oxide layer of $Al_2O_3$ is deposited and a 2D-TMD channel of monolayer $MoS_2$ is grown over it. Monolayer $MoS_2$ undergoes bandgap change caused by the transfer of polarization-induced strain from PE to TMD. We select $MoS_2$ due to its high coefficient of bandgap change for applied pressure, $\alpha_{TMD}=$ 800meV/GPa (Peña-Álvarez et al., 2015).

PE stores binary bit information (1 or 0) in the form of stable polarization states (+*P* or -*P*) respectively. The polarization state is controlled by voltage at the write port or gate to back voltage ($V_{GB}$) as illustrated by Figure 1(C)-(H). To write +*P* (logic 1), we apply $V_{GB} = V_{DD} > V_C$, where $V_C$ is the coercive voltage of PZT-5H (Figure 1(C)). $V_{GB} > V_C$ induced +*P* switching is shown by the polarization-electric field (*P-E*) response in Figure 1(D). On the contrary, application of $V_{GB} < -V_C$ causes polarization to switch to -*P* state (or logic 0) as signified in Figure 1(F, G). At a structural level, a perovskite material such as PZT-5H exist in +*P* (or -*P*) polarized state due to upward (or downward) displacement of $Ti^{4+}/Zr^{4+}$ from their centro-symmetric position, as depicted in Figure 1(E, H).

To read the stored polarization in FE, we apply a positive voltage ($V_R$) across *G* and *B*. We present a description of the read mechanism in PeFET through Figure 2. First, $V_R < |V_C|$ is applied to ensure that current state of polarization in PE is not disturbed. $V_R$ has the following role: (i) it actuates strain (piezoelectric effect) in the PE, which is in turn transduced to the TMD channel and (ii) simultaneously turns on the TMD channel. If +*P* had been stored in the PE, $V_R$ enhances charge separation along the direction of polarized charge as shown in Figure 2(A). This causes an increase in PE thickness ($\Delta t_{PE} > 0$) and yields positive strain ($S_{PE} = \frac{\Delta t_{PE}}{t_{PE}} > 0$). The experimentally characterized strain-electric field (*S-E*) response of PZT-5H reported by (Malakooti and Sodano, 2013) reflects this effect (see the *S-E* plot in Figure 2(A)). As pointed by the arrow, PZT-5H in +*P* demonstrates positive strain on experiencing a voltage that is positive but lower than $V_C$ (similar to $V_R$). In case of -*P*, $V_R$ being opposite in polarity compared to the stored polarization, diminishes charge separation (Figure 2(B)). This constricts PE thickness ($\Delta t_{PE} < 0$) resulting in negative strain ($S_{PE} < 0$, as also highlighted in the *S-E* plot of Figure 2(B)). Strain in PE translates to stress ($\sigma_{PE}$) which is induced as pressure in TMD ($\sigma_{TMD}$) and is responsible for dynamic modulation of bandgap in TMD ($\Delta E_G$). Positive strain in PE ($S_{PE} > 0$) transduces as positive pressure in ($\sigma_{TMD} > 0$) causing bandgap reduction ($\Delta E_G > 0$). Contrarily, negative strain expands the bandgap ($\Delta E_G < 0$). Note, even for $S_{PE} = 0$, TMD can experience stress from components in the device structure other than that due to the piezoelectric effect, leading the bandgap reduction from its intrinsic value. While positive $S_{PE}$ further reduces the bandgap, negative $S_{PE}$ relaxes this pressure leading to bandgap expansion towards the intrinsic value. The effect of reduced/expanded bandgap change reflects in drain current as low/high resistance states (LRS/HRS) respectively. Hence, enhanced drain to source current ($I_{DS} = I_{LRS}$) is sensed for +*P* and $I_{DS} = I_{HRS}$ is for -*P* during read.

## 2.2 Modeling and Simulation

To perform circuit simulations of PeFET, we employ a simulation framework that integrates HSPICE, COMSOL and Verilog A based models of various components in PeFETs. A representation of the modeling framework is provided in Figure 3. First, we discuss the HSPICE based circuit-compatible model Miller model used for capturing the ferroelectric behavior of PE. The equivalent circuit of the PE is shown in Figure 3. We utilize Equations (1)-(2) to simulate the polarization-electric field switching behavior of PE. Figure 4(A) presents calibration of the simulated *P-E* characteristics with experimental characterization of PZT-5H by (Malakooti and Sodano, 2013). The hysteresis window of P-E response of PZT-5H is 18kV/cm with $E_C$ = 9kV/cm (Malakooti and Sodano, 2013). The calibrated values of saturation polarization ($P_S$), remnant polarization ($P_R$), coercive electric field ($E_C$) and dielectric permittivity ($\epsilon_{r,PE}$) used in our model are provided in Table 1. The polarization switching delay, $\tau_{PE}$, is incorporated using a resistor ($R_{PE}$) – capacitor ($C_{PE}$) network (Figure 3) wherein $R_{PE} = \tau_{PE}/C_{PE}$ and $C_{PE}$ is given by Equation (3). For thickness of PE used in this work ($t_{PE}$ = 600nm), $V_C = E_C \times t_{PE}$ ~ 0.54V. Based on this, we select the write voltage of PeFET to be $V_{GB}$ = 0.8V > $V_C$. We use $\tau_{PE}$ = 1.8ns as reported by (Larsen et al., 1991) for PZT.

$$P = P_S \tanh\left(\frac{E \pm E_C}{2\delta}\right) + \epsilon_0 \epsilon_{r,PE} E \tag{1}$$

$$\delta = \alpha \left[\ln\left(\frac{P_S + P_R}{P_S - P_R}\right)\right]^{-1} \tag{2}$$

$$C_{PE} = A_{PE}\left(\frac{dP}{dE}\right) \tag{3}$$

Next, we model a 3D structure of PeFET in COMSOL Multiphysics Suite (Figure 3) that integrates solid mechanics, electrostatics and their couplings using Equations (4)-(7). Using this model, we analyze piezoelectric effect in PE and transduction of stress to 2D-TMD during read. We employ strain-charge equations (4)-(5) to our 100nm×180nm×600nm PE composed of PZT-5H. To obtain strain in PE ($S_{PE}$), we provide $V_R$ = 0.4V to the gate contact (labelled as 7 in Figure 4(B)). Therefore, $E$ across PE = $V_R/t_{PE}$ = 6.7kV/cm. $E$ translates to strain by means of piezoelectric coupling coefficients, $d$. We use parameter values of $d$ ($d_{33}$ and $d_{31}$) that are reported in (Malakooti and Sodano, 2013) based on experimentally characterized strain vs. electric field response of PZT-5H. Stress in PE $\sigma_{PE}$ (Equation (4)), generated due to interactions of various materials in the model (Equation (7)), contribute to $S_{PE}$ by means of the compliance parameter, $s_E$. Electric displacement field, $D$, caused by $\sigma_{PE}$ and $E$ is modeled using Equation (5).

$$S_{PE} = s_E \sigma_{PE} + d^T E \tag{4}$$

$$D = d\sigma_{PE} + \epsilon_0 \epsilon_{r,PE}^T E \tag{5}$$

$$\nabla \cdot D = \rho \tag{6}$$

$$\nabla \cdot \sigma_{PE} = 0 \tag{7}$$

Further, to boost efficiency of transduction of stress from PE ($\sigma_{PE}$) to TMD ($\sigma_{TMD}$), we incorporate hammer and nail effect. Hammer and nail is effective when the area of nail/2D-TMD ($A_{TMD}$) is sufficiently smaller than that of PE ($A_{PE}$), i.e., $A_{TMD} < A_{PE}$. Smaller $A_{TMD}$ than $A_{PE}$ allows stress from PE (hammer labelled as 8 in Figure 4(B)) to be better localized to TMD that lies above the nail (label 3, 7 in Figure 4(B)), thereby facilitating efficient transfer. We define a device parameter $\kappa$ in Equations (8)–(9) to help us later analysis of this principle.

$$\kappa = \frac{A_{TMD}}{A_{PE}} \tag{8}$$

$$= \frac{L_{TMD} W_{TMD}}{L_{PE} W_{PE}} < 1 \tag{9}$$

Here, $L_{TMD}$ = 20nm is the feature size of PeFET. $W_{TMD}$ is the width of TMD. We use minimum width of TMD as per design rules, $W_{PE} = W_{TMD} = 1.5 \times L_{TMD}$ = 30nm, to maintain low $\kappa$ and maximize $\sigma_{TMD}$. We choose a wide PE ($W_{PE}$) while leveraging the total device length of PeFET including contacts for $L_{PE}$. Such a design consideration allows us to achieve $L_{PE}$ = 100nm > $L_{TMD}$ that assists in further diminishing $\kappa$, without incurring additional overhead. Details about $W_{PE}$ are provided in Section 3.1.

Moreover, we choose metals with high stiffness (e.g., Pd, Cr) for the gate (beneath the nail), bottom contact of PE and source/drain contacts (Figure 4(B)). We surround the PeFET including the source/drain contacts and TMD with an encapsulant material that has high elastic modulus (e.g., $Al_2O_3$) (Schulman Daniel S., 2019). The purpose of the capping layer is to restrain the expansion of the whole PE/gate stack/TMD structure (Newns et al., 2012; Schulman Daniel S., 2019). By constraining the TMD from the top, it helps to localize the piezoelectricity induced strain in PE towards compressing the TMD material (via by the gate stack).

We use $\sigma_{TMD}$ obtained from the COMSOL model as input to Verilog-A model of 2D-TMD FET. This model first converts $\sigma_{TMD}$ to bandgap change, $\Delta E_G = \alpha_{TMD}\sigma_{TMD}$, where $\alpha$ is the bandgap coefficient of TMD (Table 1). We use a capacitive network-based model (Suryavanshi and Pop, 2016) modified for a back gated device to model the electrostatics of the 2D FET. The charge density and source/drain quasi-Fermi level of a TMD material are self-consistently solved in the model (Suryavanshi and Pop, 2016). We incorporate the effect of bandgap modulation ($\Delta E_G$) induced by transduction of piezoelectric strain (Equation 10) in the calculation for quasi-Fermi level. Next, the continuity equation is used to derive drain to source current of TMD (Suryavanshi and Pop, 2016). The drain to source current (Equation 11) reflects not only the effect of electrostatics but also that of bandgap modulation in PeFET device characteristics.

$$E_G = E_0 - \Delta E_G \tag{10}$$

$$I_{DS} = f(E_G, V_{GS}, V_{DS}) \tag{11}$$

where, $E_0$ is the bandgap of TMD at zero gate to back voltage.

Finally, the HSPICE compatible model of PeFET is a combination of Miller equation for PE/FE with polarization induced piezoelectric response incorporated 2D-TMD FET model. The parameters used in our simulations are based on prior literature and experiments (Table 1).

## 3 Characteristics of 2D Piezoelectric FET

### 3.1 Strain Transfer through Hammer-and-Nail Principle

To analyze the hammer and nail principle in our 3D COMSOL model of PeFET (Figure 4(B)), we use $W_{PE} = 180$nm, that results in $\kappa = 0.03 < 1$ according to Equation (9). We show ~11× increase in $\sigma_{TMD}$ compared to $\sigma_{PE}$ for $V_R = V_{GB} = 0.4$V in Figure 4(B). At this $V_R$, $\sigma_{TMD}$ causes bandgap of TMD to decrease (increase) by 48.4mV when PE is in +P/-P state.

Tuning of $\kappa$ enables design time optimization of the distinguishability of memory states in PeFET. We know from Section 2.1 that positive stress appears in a +P polarized PeFET on application of $V_{GB} = V_R$. By decreasing $\kappa$, we further enhance the hammer and nail effect or localization of positive stress on TMD. As a result, resistance of TMD decreases to a greater extent. Hence, $I_{LRS}$ increases. Contrarily, for -P, negative stress caused by $V_R$ is accentuated for smaller $\kappa$. This results in a more resistive HRS in TMD ($I_{HRS}$ decreases). The combined effect of improved $I_{LRS}$ and diminished $I_{HRS}$ improves distinguishability (= $I_{LRS}/I_{HRS}$) significantly. According to our approach in Section 2.2, we increase $W_{PE}$, keeping other dimensions fixed, to achieve lower $\kappa$. This leads to a tradeoff between improved distinguishability and area increase which, in turn, can potentially increase latency and energy. Considering these aspects, we design our PeFET here with $\kappa = 0.03$ that provides us with a distinguishability of 5× (details in next section) and sufficient drain current for desirable sense margin for dot product computations (elaborate discussion in Section 5.3).

### 3.2 Device Characteristics of PeFET

Let us start with a brief discussion on the biases required for $\pm P$ storage in PeFET. To write $+P$ (or 1), we provide $V_{GB}$ with $0.8V = V_{DD} > V_C$ of PZT-5H (= 0.54V at $t_{PE}$ of 600nm as per Figure 4(A)). Similarly, $-P$ (0) is stored at $V_{GB} = -0.8V < -V_C$.

Now, we divulge into the polarization/strain-dependent transfer characteristics ($I_{DS}$-$V_{GS}$) of PeFET. To avoid polarization-switching while obtaining transfer characteristics, we apply a positive gate voltage $V_G = V_R = 0.4V$ ($< V_C$ of PZT-5H = 0.54V at $t_{PE}$ = 600nm) while the back contact ($V_B$) is kept at 0V akin to Figure 5(A). Note that $V_R$ at the gate turns on the channel (controls electrostatics) while triggering piezoelectric response by dint of $V_{GB} = V_R$ across PE. For comparison, we also simulate a device with $V_{GB} = 0$ (sweeping $V_G$ and $V_B$ at the time), from which we obtain polarization-independent nominal transfer characteristics of MoS$_2$ based 2D FET.

When $V_{GB} = V_R = 0.4V$ is applied, PeFET with $+P$ undergoes positive strain in PE (follow grey arrow in Figure 5(B)) that results in bandgap reduction $\Delta E_G = 48.4$mV, yielding 2.3× enhanced $I_{DS}$ (= $I_{LRS}$) compared to the baseline (MoS$_2$-FET with $V_{GB} = 0$), as shown in Figure 5(C). Contrarily, when a $-P$ state PeFET receives the same $V_R$, $I_{DS}$ diminishes by 2.2× compared to baseline which we refer to as $I_{HRS}$ (Figure 5(C)). This is because of negative strain (follow orange arrow in Figure 5(B)) in PE, which ultimately reflects as increase of bandgap towards the intrinsic value. Note that these results correspond to $V_{DS} = 0.8V$ and $\kappa = 0.03$. Overall, the distinguishability or $I_{LRS}/I_{HRS} =$ ~5×.

Let us now present the PiER characteristics of PeFETs, which is associated with the dependence of PeFET characteristics on the polarity of $V_{GB}$ and eventually enables us to design signed ternary CiM.

### 3.3 Polarization Preserved <u>Pi</u>ezoelectric <u>E</u>ffect <u>R</u>eversal with Dual Voltage Polarity (PiER)

Until now, our analyses focused on piezoelectric response generated when PE is subjected to $V_{GB} = V_R > 0$. Recall that we maintain $V_B = 0$, while sweeping $V_G$ to $V_R$ to achieve the same. With this bias, PeFET in $+P$ yields LRS whereas $-P$ leads to HRS (Figure 5(C)).

Interestingly, the sensed resistance states with $\pm P$ are reversed when voltage across PE is negative, i.e., $V_{GB} = -V_R < 0$. Again, since $V_R < |V_C|$, stored state of polarization is undisturbed. For the same polarization stored in PE, negative $V_{GB}$ induces opposite piezoelectric response in PE compared to positive $V_{GB}$. This allows the same polarization to induce opposite resistance states in TMD for $V_{GB} = -V_R$ compared to $V_{GB} = V_R$. Note that we bias $V_G = V_R = V_{DD}/2$ and $V_B = V_{DD}$ respectively such that $V_{GB} = -V_{DD}/2 = -V_R$, also illustrated in Figure 5(D). Since $V_G$ controls electrostatics in TMD apart from piezoelectricity in PE, we ensure that a positive gate voltage greater than the device threshold voltage is applied, to keep the PeFET ON even when $V_{GB} < 0$.

We elucidate the reversal of piezoelectric effect and its impact on the TMD resistance now. Let the stored polarization in PE be $-P$. When $V_{GB} = -V_R$, charge separation occurs in the same direction as that of initial polarization. This causes $t_{PE}$ to elongate, thereby generating positive strain in PE for $-P$ (follow grey arrow in Figure 5(E)). We know from our previous understanding that bandgap reduction of TMD occurs when it receives positive strain. Hence, PeFET is in LRS for $-P$ state. For $+P$, negative $V_{GB}$ (= $-V_R$) bias reduces the polarization. This causes $t_{PE}$ to constrict, negative strain (orange arrow in Figure 5(E)) is generated that leads to bandgap increase. Hence, HRS is observed for $+P$. Thus, <u>$+P$ ($-P$)</u> stored in PeFETs can be configured in a <u>low (high)</u> resistance state by applying $V_{GB} > 0$ (i.e. $V_B = 0$, $V_G = V_R = V_{DD}/2$). On the other hand, <u>$+P$ ($-P$)</u> induces <u>high (low)</u> resistance states with $V_{GB} < 0$ (i.e. $V_B = V_{DD}$, $V_G = V_R = V_{DD}/2$) during sensing (read/compute). We name this unique property as Polarization Preserved <u>Pi</u>ezoelectric <u>E</u>ffect <u>R</u>eversal with Dual Voltage Polarity (PiER). We will refer to piezoelectric effect pertaining to $V_{GB} = -V_R$ as *PiER<u>Ce</u>*, where Ce signifies that negative $V_{GB}$ mode is used *exclusively* for ternary <u>c</u>ompute (Section 5). Whereas, *PiER<u>Re</u>* identifies piezoelectric effect for positive $V_{GB}$ (used for standard <u>re</u>ad as well as compute). We summarize this discussion in Table 2.

From our analysis of PeFET device characteristics in *PiERCe* configuration (Figure 5(F): $V_G = 0.4V$, $V_B = 0.8V$ and $V_{GB} = -0.4V$), we observe that PeFET with $-P$ exhibit 2.3× larger drain current ($I_{LRS}$) whereas that with +P shows 2.2× lower drain current ($I_{HRS}$) compared to baseline (i.e. PeFET without bandgap modulation: $V_{GB}$ = 0). Overall, distinguishability = ~5× is achieved, which is similar to that for read described in Section 3.2, but with polarization state mapping to LRS and HRS swapped.

Note that we use a constant electron mobility, $\mu_e$ = 90cm$^2$/Vs for MoS$_2$ in our PeFET model (Hosseini et al., 2015; Yu et al., 2017). However, studies have shown that mobility of MoS$_2$ improves (degrades) subject to positive (negative) uniaxial strain such as that experienced by PeFET (Hosseini et al., 2015). Note that in PeFET, LRS and HRS are outcome of positive and negative strain respectively. This implies improvement of $I_{LRS}$ (due to enhanced $\mu_e$) and degradation of $I_{HRS}$ (caused by lowered $\mu_e$). Consequently, a higher distinguishability of PeFET may be expected than the reported value in this work.

## 4    Ternary Compute Enabled Memory with PeFET

In this section, we propose a PeFET based non-volatile memory with the capability to perform dot product computations in the signed ternary regime. We refer to the proposed memory as Strain-enabled Ternary Precision Computation-in-Memory (STeP-CiM).

### 4.1    STeP-CiM Cell

STeP-CiM presented in Figure 6(A, B) consists of two PeFET based bit cells ($M_1$ and $M_2$). $M_1$ and $M_2$ store bit information (1/0) in the form of +P/-P polarization. $M_1$ and $M_2$ use 2D-TMD FET based access transistors ($AX_1$, $AX_2$, $RAX_1$, $RAX_2$) that are switched on/off using word line (WL). Access transistors $AX_1$, $AX_2$ connect bit lines $BL_1$ and $BL_2$ with the gate terminals ($G_1$, $G_2$) of the respective PeFETs $M_1$ and $M_2$. Recall that the gate terminal is a common control knob for the channel of the 2D-TMD FET and PE in $M_1$/$M_2$. Hence, $BL_1$ and $BL_2$ can actuate ferroelectric switching for write as well as piezoelectric response in PE for read/compute depending on the voltage they are driven to. The bias conditions of $BL_1$/$BL_2$ and impact on write-read-compute operation are discussed in detail in Section 4.2 and 5. Note that $RAX_1$ and $RAX_2$ are read access transistors that connect drains ($D_1$, $D_2$) of PeFETs in $M_1$ and $M_2$ to read bit lines $RBL_1$ and $RBL_2$, respectively. The back terminals of PeFETs in $M_1$ and $M_2$ are shared, and connected to compute word line, CWL. Read and compute are achieved by sensing strain-induced resistance changes in the PeFETs (more in Sections 4.2.2, 5) in terms of $RBL_1$ and $RBL_2$ currents. During hold, voltages of $BL_1$, $BL_2$, $RBL_1$, $RBL_2$, CWL and WL are 0V.

Note that $M_1$/$M_2$ of STeP-CiM cell can be used as standard memory with binary storage. Hence, STeP-CiM cell can be reconfigured to serve as a standard memory (with 2 bit-cells) or a compute-enabled memory for ternary precision as per application needs (further discussion on this in Section 6). Using two access transistors (such as $AX_1$ and $RAX_1$ in $M_1$) does not lead to any area penalty in the layout shown in Figure 6(C). This is because the layout area is dictated by the PeFET footprint arising from the wide PE requirement for hammer and nail effect. As per our layout analysis, both AX and RAX can be accommodated within the PE area.

The access transistors in STeP-CiM cell ($AX_1$, $AX_2$, $RAX_1$, $RAX_2$) serve two other purposes, in addition to achieving selective access to the cells in a memory array. First, $AX_1$/$AX_2$ of the un-accessed cells disconnect $BL_{1/2}$ from the respective PE capacitance, which is large due to high dielectric permittivity of PZT-5H, $\epsilon_{r,PE}$ = 4000 (Malakooti and Sodano, 2013). This averts the increase in the total BL capacitance due to large PE capacitance ($C_{PE}$) and improves write energy efficiency and performance. Second, $RAX_1$/$RAX_2$ provides means to disconnect un-accessed PeFET from RBLs, thereby avoiding unwanted RBL currents. It is an important aspect in this design as floating gate terminals of PeFETs in the un-accessed cells (disconnected from $BL_{1/2}$ by $AX_1$/$AX_2$) may develop a potential greater than the threshold voltage of TMD FET due to noise and leakage, leading to spurious currents on RBLs. The advantage of using an additional access transistor is decoupling of write and read/compute operations, which enhances the design margins, especially for the dot product computation. With this background, we now describe the write and read operations next, and CiM operations in the subsequent section.

## 4.2 Write and Read Operations of STeP-CiM cell

### 4.2.1 Write

The encoding for signed ternary weights stored in a STeP-CiM cell is provided in Table 3(A). To store ternary '1' in STeP-CiM, $+P$ and $-P$ are written in $M_1$ and $M_2$ as per Table 3(A). This operation which is depicted by Figure 6(D)-(F). First, $BL_1$ is driven to $V_{DD} > V_C$ and $BL_2$ to 0V. $RBL_{1/2}$ are kept at 0V. Next, $WL$ is asserted to $V_{DD}+V_{TH}$ (boosted to compensate for threshold voltage $V_{TH}$ drop in write access transistors). Finally, $CWL$ is supplied with a two-phase signal $(0 \rightarrow V_{DD})$, wherein the voltage in the first phase $(\Phi_1)$ is 0V, while it is $V_{DD}$ in the second phase $(\Phi_2)$. The two-phase signal (Li et al., 2019) facilitates writing '1' and '0' states to multiple PeFETs as follows. PeFET in $M_1$ (Figure 6(E)) experiences $V_{GB} = V_{BL1} - V_{CWL} = 0.8V$ during $\Phi_1$ since $V_{BL1} = 0.8V$ and $V_{CWL} = 0V$. This results in $-P \rightarrow +P$ switching. $M_2$ (Figure 6(F)) experiences $V_{GB} = V_{BL2} - V_{CWL} = 0$ (as $V_{BL2} = 0V$ and $V_{CWL} = 0V$) during $\Phi_1$ and the previous polarization state is preserved. During $\Phi_2$, $M_1$ retains its state of $\Phi_1$ ($V_{GB} = 0$) while $M_2$ switches to $-P$ after receiving $V_{GB} = -0.8V$ ($V_{BL2} = 0V$ and $V_{CWL} = 0.8V$). Similarly, for ternary '-1', $-P$ and $+P$ should be written to $M_1$ and $M_2$ (Table 3(A)). The process is similar except that now, $BL_1$ is driven to 0V and $BL_2$ to 0.8V. Finally, ternary '0' corresponds to $-P$ in both $M_1$ and $M_2$, which is stored by having $BL_1$ and $BL_2$ at 0V.

### 4.2.2 Read

In order to sense the stored polarization value in the STeP-CiM cell, a *positive* $V_{GB}$ $(= V_R < V_C)$ need to be applied across the PEs of $M_1$ and $M_2$ for them to be in *PiERRe* condition (refer to Table 2). Moreover, gates $G_1$ and $G_2$ of $M_1$ and $M_2$ should receive $V_R$ for PeFETs to conduct. To achieve this, we drive $BL_1$ and $BL_2$ to $V_R = 0.4V$ while $CWL$ is kept at 0V. In addition, $RBL_1$ and $RBL_2$ are switched to $V_{DD} = 0.8V$ to facilitate drain to source conduction of PeFETs. The schematic with biases for the read operation and waveform are demonstrated in Figure 7(A, B).

On asserting $WL$ with $V_{DD} = 0.8V$, $V_{GB} = V_{BL1/BL2} - V_{CWL} = 0.4V$ for both $M_1$ and $M_2$. Let us explore the sensing of ternary '1'. In this case, as $+P$ is stored in $M_1$, bandgap reduces $(\Delta E_G < 0)$ in response to $V_{GB} = 0.4V$ and $I_{LRS}$ is sensed on $RBL_1$ (corroborating with Table 2). Contrarily, for $-P$ in $M_2$, bandgap expands with $V_R$, i.e., $\Delta E_G > 0$ (dotted line of Figure 7(A)) leading to increased resistance of $M_2$ or $I_{HRS}$ on $RBL_2$. $I_{LRS}$ on $RBL_1$ and $I_{HRS}$ on $RBL_2$ indicate ternary '1' storage, as also listed in Table 3(B). For ternary '-1' ( $-P$ in $M_1$ and $+P$ in $M_2$), we obtain $I_{HRS}$ on $RBL_1$ and $I_{LRS}$ on $RBL_2$. For ternary '0' which is encoded by $-P$ in both $M_1$ and $M_2$, $I_{HRS}$ is observed on $RBL_1$ and $RBL_2$.

## 4.3 Segmented Architecture of STeP-CiM

If standard memory array architecture is followed for STeP-CiM cell wherein $CWL$ runs throughout the row, $C_{PE}$ from all cells in the row add to $CWL$ capacitance. This could lead to large energy overheads (Thakuria and Gupta, 2022), since $C_{PE}$ for PZT-5H is large (as discussed before). To mitigate this, we design an array for STeP-CiM that employs segmentation similar to FERAMs (Rickes et al., 2002). Figure 8 illustrates the segmented array architecture of STeP-CiM based cells. Segmentation may not be required for CiM in DNNs that utilize high parallelism by computing the dot products for all the columns simultaneously. However, if this proposed array is used as a standard memory (as discussed above), segmentation will be important for high energy efficiency, especially in edge devices. Therefore, we employ the segmented architecture with an objective to support the reconfiguration of the proposed design from a compute-enabled ternary memory for DNNs to a standard memory, as per the application needs.

A segment in the segmented array (Thakuria and Gupta, 2022) is sized as 64×256 (Figure 8). Each segment has an exclusive global plate line (*GPL*) that runs along the column direction. *GPL* acts as an input to buffers in each local row of the segment. The output of the buffers is used to drive a local read word-line *LCWL* for each local row comprised of 64 STeP-CiM cells. Notice that, the capacitance on *LCWL* is from $C_{PE}$ of 64 STeP-CiM cells instead of the entire row, which enhances the energy-efficiency. *WL* provides the supply voltage to the buffers and also activates access transistors of each STeP-CiM in the accessed segment. Bit-lines $BL_1$, $BL_2$, $RBL_1$ and $RBL_2$ run along the column. The 64 STeP-CiM cells in a segmented row are accessed simultaneously for read and write.

Appropriate biasing of *GPL* during write and read operations is important to ensure that *LCWL* voltage is identical to *CWL* voltage discussed in Section 4.2. For write, we apply the two phase $0 \rightarrow V_{DD}$ signal to *GPL*, instead of *CWL* in Section 4.2.1. When *WL* is asserted with $V_{DD} + V_{TH}$, *LCWL* is driven to $0 \rightarrow V_{DD}+V_{TH}$ by the active buffers connected to *GPL* and *LCWL*. $V_{GB} = V_{DD}$ in $\Phi_1$ and $-P \rightarrow +P$ write occurs, while $+P \rightarrow -P$ occurs in $\Phi_2$ when $V_{GB} = -V_{DD}$, similar to Section 4.2.1. During read, *GPL* voltage is 0V with $WL = V_{DD}$ such that *LCWL* is at 0, as in Section 4.2.2. Other lines are biased in an identical fashion as described in Section 4.2.1 (for write) and 4.2.2 (for read). *WL* is de-asserted for all un-accessed rows of an accessed segment. A segment is put on hold by pulling its *GPL*, *WLs* (other than that of the row accessed by another segment) and all *RBLs*, *BLs* to 0V.

## 5  In Memory Ternary Computation using STeP-CiM

In this section, we explain ternary in-memory scalar multiplication and dot product computation using STeP-CiM. We target signed ternary precision for weights, inputs and the scalar product having values {-1, 0, 1} (Li et al., 2016). As discussed in Section 4.2.1, combination of polarization states of $M_1$ and $M_2$ in STeP-CiM constitute a ternary weight (Table 3(A)). The ternary inputs encoded with *WL* and *CWL* voltages to utilize the resistance states of both conditions - *PiERRe* (*CWL* = 0) and *PiERCe* (*CWL* = $V_{DD}$ = 0.8V) are indicated in Table 3(C). More details on this follow below. $BL_1$, $BL_2$ are driven to $V_R = 0.4V$ so that $V_{GB} < |V_C|$ appears across PE of $M_1$ and $M_2$ (similar to Section 4.2.2). $RBL_1$ and $RBL_2$ are driven to $V_{DD}$ during compute. In accordance with the ternary weights and applied input, different instances of $RBL_1$ and $RBL_2$ currents ($I_{RBL1}$ and $I_{RBL2}$) are observed. Finally, the scalar product or output is obtained as $O = I_{RBL1} - I_{RBL2}$. Notice from Table 3(D) that $O = \{\underline{-1}, 0, \mathbf{1}\}$ is interpreted as $\{(\underline{I_{HRS} - I_{LRS}}), 0, (\mathbf{I_{LRS} - I_{HRS}})\}$ respectively.

### 5.1  Ternary Scalar Multiplication using STeP-CiM

Before delving into details of ternary scalar multiplication with STeP-CiM, we elaborate on what the input encoding (*I*) in Table 3(C) represents in terms of resistance states. Subsequently, we evaluate examples of ternary scalar multiplication. The truth table for scalar product is available in Table 3(E).

**Ternary input (*I*) = +1:**

$I = +1$ corresponds to *CWL* = 0 and *WL* being asserted with $V_{DD}$. With $BL_1$ and $BL_2$ being $V_R$ during compute (as mentioned above), we have $V_{GB1,2} = V_{BL1,2} - V_{CWL} = V_R$ for $I = +1$. Note that $V_{GB}$ being a positive voltage here, puts PeFETs in *PiERRe* resistance regime (corroborating with Table 2). That is, $+P$ is read as LRS ($I_{LRS}$) and $-P$ as HRS ($I_{HRS}$). With this background, we elaborate the scalar products for different weight (*W*) conditions with $I = +1$ (for which PeFETs are in *PiERRe*). Please refer to Table 3(E) for further clarity on the descriptions of *W*, *I* and corresponding *O*.

(a) ***W* = +1**: According to this weight encoding, $M_1$ and $M_2$ store $+P$ and $-P$ respectively. Since, PeFETs are in *PiERRe* because of $I = +1$, $M_1$ and $M_2$ are in LRS and HRS respectively. Hence, $I_{RBL1} = I_{LRS}$, $I_{RBL2} = I_{HRS}$, and $O = W \times I = I_{LRS} - I_{HRS}$. *O* corresponds to scalar product of +1 in Table 3(E). Figure 7(A)-(B) shows the waveform for this example.

(b) *W* = -1: $M_1$ and $M_2$ are written with -P and +P respectively, hence they exhibit HRS and LRS for $I = 1$. Hence, $I_{RBL1} = I_{HRS}$ and $I_{RBL2} = I_{LRS}$ and $O = I_{HRS} - I_{LRS}$ corresponding to scalar product = -1.

(c) *W* = 0: Both $M_1$ and $M_2$ have -P stored in them, and are in HRS for $I = 1$. Thus, $I_{RBL1} = I_{HRS}$, $I_{RBL2} = I_{HRS}$, $O = I_{HRS} - I_{HRS} = 0$ (corresponding to scalar product of 0).

**Ternary input (*I*) = -1:**

For $I = -1$, *CWL* and *WL* are both switched to $V_{DD}$. Since, $BL_1$ and $BL_2$ remain at $V_R$ (= $V_{DD}/2$) during compute, we have $V_{GB1,2} = V_{BL1,2} - V_{CWL} = -V_{DD}/2 = -V_R$ for $I = -1$. With $V_{GB} < 0$, now PeFETs $M_1$ or $M_2$ are in *PiERCe* resistance regime. Hence, +P and -P are sensed as HRS ($I_{HRS}$) and LRS ($I_{LRS}$). Note that the sensed states are reversed for the same stored polarization compared to previous example due to *PiERCe* (refer to Section 3.3 for detailed mechanism). The scalar products with $I = -1$ for varying weights are evaluated below.

(a) *W* = +1: Although $M_1$ and $M_2$ have +P and -P stored in them (same as in example 5.1(a)), they now exhibit HRS and LRS respectively now due to PeFETs being in *PiERCe*. This is caused by interaction of the stored polarization with negative $V_{GB}$ (refer to Table 2) when $I = -1$. Ultimately, $I_{RBL1} = I_{HRS}$, $I_{RBL2} = I_{LRS}$ and $O = I_{HRS} - I_{LRS} = -1$ (Table 3(E)). Figure 7(C)-(D) represent this example with waveforms, highlighting the differences from $I = 1$ and $W = 1$.

(b) *W* = -1: In this case, polarization in $M_1$ and $M_2$ is -P and +P respectively. Due to *PiERCe*, $I_{RBL1} = I_{LRS}$, $I_{RBL2} = I_{HRS}$ and $O = I_{LRS} - I_{HRS} = +1$.

(c) *W* = *0*: With $M_1$ and $M_2$ both storing -P and -P. Hence, $O = I_{LRS} - I_{LRS} = 0$.

**Ternary input (*I*) = 0:**

In this case, *CWL* and *WL* are de-asserted with 0V. PeFETs are non-conducting. $I_{RBL1}$ and $I_{RBL2}$ are 0V, hence O = 0, irrespective of the weights.

### 5.2 Ternary Multiply and Accumulate (MAC) with STeP-CiM

In this section, we elaborate on the design details of a STeP-CiM array for achieving ternary MAC, with reference to the schematic in Figure 9(A). Prior to the operation, weight vector with $W_i$s is mapped and programmed to $M_{1i}$ and $M_{2i}$ of each row of STeP-CiM, following the procedure discussed in Section 4.2.1. The input vector ($I_i$) encoded as *WL* and *CWL* voltages is applied to the rows accessed for MAC. Currents flowing through $RBL_1$ and $RBL_2$ due to scalar product of $I_i$ and $W_i$ add up on the respective lines. These currents are used to evaluate the dot-product. Our method for current based sensing is as follows: first, we compare $I_{RBL1}$ and $I_{RBL2}$ to determine which branch has higher current. The output of the comparator in Figure 9(B) determines the sign ($Sn$) of the final MAC output. If $I_{RBL1} > I_{RBL2}$, $Sn = 1$ whereas for $I_{RBL1} < I_{RBL2}$, $Sn = -1$. Next, the comparator output is fed to a current subtractor circuit (Figure 9(C)) which determines the magnitude of the difference of bit currents, $I_{RBL1}$-$I_{RBL2}$. The output of the subtractor is actually an integer multiple of $I_{LRS} - I_{HRS}$, i.e., $I_{RBL1}$-$I_{RBL2} = a(I_{LRS} - I_{HRS})$, where '*a*' is the integer multiple. To determine the value of '*a*', we employ a flash Analog to Digital Converter (ADC) such as in Figure 9(D). Finally, the dot product is computed as $O = Sn \times a = \pm a$ depending on which of $I_{RBL1}$ and $I_{RBL2}$ is greater as discussed earlier. Notice that our method of subtracting of *RBL* currents before digitization of the sensed current from the array saves us an ADC compared to other ternary designs that employ ADCs on each bit line (Jain et al., 2020; Thirumala et al., 2020) due to their use of voltage-based sensing. The benefits of this are evidenced at the system level results.

Next, we throw light on the design of our peripherals and the non-idealities caused by their interaction with current based sensing scheme for MAC. The read-bit line drivers in Figure 9(A, B) used for biasing $RBL_1$ and $RBL_2$ to $V_{DD}$ during MAC operation (as per the biasing scheme discussed in Section 5) are the primary source of non-idealities. Note in Figure 9(B) that the transistor $P_{11}$ ($P_{21}$) of the comparator is connected in series to transistor $P_{12}$ ($P_{22}$) of read bit line driver, with drain of $P_{12}$ ($P_{22}$) connected to $RBL_1$ ($RBL_2$). Although this configuration is necessary for mirroring $RBL_1$ and $RBL_2$ current to the comparator required for MAC (whose functionality we have discussed previously), rising current on $RBL_1$ ($RBL_2$) with multiple row access causes voltage on the source node $S_1$ (or $S_2$) of $P_{12}$ ($P_{22}$) to be pulled to value less than $V_{DD}$ by resistive divider action of the pull up transistors of comparator/read bit-line and access transistors on $RBL$. This leads to non-ideal current on $RBL_1$ and $RBL_2$. We reference this as loading effect in the future. In other words, $RBL_{1/2}$ are biased at a value less than $V_{DD}$ due to the loading effect, and this value is dependent on $RBL$ current. Higher the $RBL$ current, larger is the voltage drop across the biasing transistors, and lower is the $RBL$ voltage. In our analysis presented in the subsequent section, we discuss the loading effect for STeP-CiM array and how it can alter the sense margin from one output to another, which is an undesirable effect.

Before proceeding to investigate the sense margin for different outputs, it is important to reflect on the number of cells that can be accessed together robustly while performing the MAC operation. We decide the same on the basis of ADC precision and sparsity of input and weight vectors. Higher ADC precision has been shown to overshadow energy efficiency achieved at the array level with CiM (Jain et al., 2020). Therefore, following their energy estimations, we consider the 3-bit flash ADC of Figure 9(D). Moreover, DNNs are known to exhibit > 50% sparsity. Considering this into account, we assert $N_V = 16$ cells simultaneously to obtain a maximum dot product output of 8, which can also be robustly computed by the 3-bit ADC. This analysis and the design decisions have been borrowed from our earlier work on ternary memories (Jain et al., 2020; Thirumala et al., 2020). It is noteworthy that outputs > 8 (rare due to sparsity > 50%) are interpreted as 8 by the system (due to limited ADC precision). However, this has negligible impact on the overall system accuracy, as confirmed by our system analysis described later.

## 5.3 Sense margin and variation analysis for Signed Ternary MAC

We evaluate the robustness of signed ternary MAC operation performed in a column of 16 rows. We study different instances of accessing word-lines 1…16 to understand their effect on $RBLs$ loading and its translation to sense margin. In essence, we want to establish combinations of $I_i$ and $W_i$ that reflect minimum loading (best case) and maximum loading (worst case) of $RBLs$ to define sense margin.

(A) Let us first consider the case where the loading effect is minimum (i.e. with lowest RBL current). To start with, we first analyze the condition for scalar product, $O = 1$. Corroborating with our previous understanding of scalar product computation in Section 5.1, we expect $I_{LRS}$ on $RBL_1$ and $I_{HRS}$ on $RBL_2$ for this output. We provide an input sequence where a row (say row$_1$) receives $I_1 = 1$ and the remaining 15 rows (e.g., rows$_{2…16}$) receive $I_{2…16} = 0$. This is achieved with $W_1 = 1$ for $I_1 = 1$. Rows$_{2…16}$ do not contribute significantly to currents on $RBLs$ as $I_{2…16} = 0$ ($WL = 0$V, which disconnects PeFETs from $RBLs$). Similarly, to obtain a MAC output of '$a$', '$a$' number of rows store $W_{1…a} = 1$ and receive $I_{1…a} = 1$. The remaining rows receive input, $I_{a+1…16} = 0$. $W$s of rows a..16 do not matter since they are non-contributing by dint of $I = 0$. Hence, $I_{RBL1} = aI_{LRS}$, $I_{RBL2} = aI_{HRS}$ and $O_a = a(I_{LRS} - I_{HRS}) = a$. Here, $a = $ number of rows with $I = 1$ and $a \leq 16$. Note that the $RBLs$ in this example are loaded with currents only from the rows having $I = 1$, which is akin to a scenario of minimum loading of $RBL$ for a desired output. This example is illustrated in Figure 10(A).

(B) Next, we consider another example whose expected outcome is similar to the case study in (A), but with $W_i$ and $I_i$ different from example (A). Here, our intent is to obtain the combinations of $W_i$ and $I_i$ that maximizes current on $RBLs$ to mimic a worst-case example of loading effect. Again, starting with $O = 1$, we program the weight of row$_1$ as $W_1 = 1$ (i.e., $M_1$:+P, $M_2$: -P) and remaining rows$_{2…16}$ with $W_{2…16} = 0$ (their $M_1$: -P, $M_2$: -P). The inputs corresponding to row$_1$ $I_1 = 1$ and rows$_{2…16}$ $I_{2…16} = -1$.

We expect a MAC output = 1 using these combinations. Let us analyze what this means in terms of scalar product from each row, and the resultant MAC output.

The cell in row$_1$ with $I_1 = 1$ is in *PiERRe* mode. This implies that for $W_1 = 1$, $M_1$ is in LRS and $M_2$ in HRS. Correspondingly, the contribution to $I_{RBL1}$ and $I_{RBL2}$ is $I_{LRS}$ and $I_{HRS}$. Rows$_{2...16}$ with $I = -1$ are in *PiERCe*. Hence, for $W_{2...16} = 0$ (-P,-P as per Table 3(A)), both $M_1$ and $M_2$ are in LRS (Table 2), we observe $I_{LRS}$ on $RBL_1$ and $RBL_2$. Ultimately, we obtain $I_{RBL1} = 16I_{LRS}$ and $I_{RBL2} = I_{HRS}+15I_{LRS}$. Overall, $O_1 = I_{RBL1} - I_{RBL2} = I_{LRS} - I_{HRS}$, which corresponds to output of 1. However, $I_{RBL1}$ and $I_{RBL2}$ in this scenario is significantly higher than example (A), reflecting worst case loading effect.

Similarly, to obtain a MAC output of '$a$' while loading the $RBLs$ maximally, '$a$' number of rows get input and weight as 1 (i.e, $I_{1...a} = 1$, $W_{1...a} = 1$) which contribute as $I_{RBL1} = aI_{LRS}$ and $I_{RBL2} = aI_{HRS}$. The remaining rows receive input of -1 and weight 0 (i.e., $I_{a+1...16} = -1$, $W_{a+1...16} = 0$). Hence, from these rows we receive $I_{RBL1} = (16-a)I_{LRS}$ and $I_{RBL2} = (16-a)I_{LRS}$. For all the 16 rows, $I_{RBL1} = 16I_{LRS}$ and $I_{RBL2} = 16I_{LRS} + a(I_{HRS} - I_{LRS})$ and $O_a = a(I_{LRS} - I_{HRS}) = a$.

From (A) and (B), it is clear that the former and latter have highest and lowest loading effects. Therefore, they set the case for maximum and minimum currents for expected outputs (Figure 10(A, B)). Based on this approach, we define the worst case sense margin for an expected output '$a$' (say) to be $= (O_{Min\_load,a} - O_{Max\_load,a-1})/2$. Here, $O_{Min\_load,a}$ is based on minimal loading of $RBL_1$ and $RBL_2$ for output '$a$' calculated using the method in (A), while $O_{Max\_load,a-1}$ is the maximum loading of $RBL_1$ and $RBL_2$ for the prior output '$a-1$' using method in (B). Figure 10(B) depicts this method of calculating sense margin. The calculated sense margin is plotted in Figure 10(C). Note that the minimum sense margin of $> 1\mu A$ is obtained by optimizing the widths of the loading transistors in the read bit-line drivers.

We further perform variation analysis (Figure 10(E)) using Monte Carlo HSPICE simulations and analyze the sensing errors in ternary MAC based on sense margins in Figure 10(D). We consider $\sigma = 10mV$ random variation of $V_{TH}$ (Smets et al., 2019; Sebastian et al., 2021) in transistors in STeP-CiM. As the expected MAC output increases, we observe overlap of output currents with adjacent states resulting in an error magnitude of $\pm 1$ and rising trend of sensing error probability. We calculate a total of such 10 errors from 16 outputs, each undergoing 1000 Monte Carlo iterations. Combined with occurrence probability of error for each state (Jain et al., 2020), the overall error is expected to not affect DNN accuracy.

## 5.4 Architecture for increased parallel computation of MAC

Next, we discuss the STeP-CiM array used for performing parallel in-memory dot product computation between ternary inputs and weights. The size of our STeP-CiM array is 256×256 (= $N_R \times N_C$). The array is segmented into 16 blocks, wherein each block consists of 16×256 (= $N_V \times N_C$) STeP-CiM cells. All $N_V$ rows and $N_C$ columns of the block are asserted during a block access for dot product computation. Hence, a block can perform simultaneous ternary multiplication of input vector $I$ with $N_V$ elements and weight matrix $W$ of size $N_V \times N_C$. We follow a similar architecture as proposed in (Jain et al., 2020) to compute dot-product with input vectors $N_V > 16$. In this case, partial sums are stored in a peripheral compute unit (PCU) using a sample and hold circuitry. The partial sums are accumulated after several block accesses to get the final dot product. The dot products are then quantized, and passed through an activation function to provide inputs to the next DNN layer (Jain et al., 2020). We use $Q = 32$ PCUs for the entire array (where $Q < N_C = 256$) to minimize area/energy overheads of the peripheral circuits (Jain et al., 2020).

## 6 Results and Analysis

## 6.1 Array-Level Analysis

Here, we present analysis of STeP-CiM for array level metrics, namely cell area, latency and energy for write, read and MAC operations. We compare them with near-memory designs based on PeFETs (PeFET-NM) and 2D FET based SRAM (SRAM-NM). The STeP-CiM cell presented in Figure 6(A) can be readily repurposed for near memory compute by maintaining $CWL = 0V$ (akin to *PiERRe* condition), during these operations. We name this mode as PeFET-NM. Whereas, during in-memory ternary dot-product computations, STeP-CiM operate with either $CWL = 0$ (*PiERRe*) for $I = 1$ or $CWL = V_{DD}$ (*PiERCe*) for $I = -1$. SRAM-NM cell is designed with two 2D FET SRAM bit cells. The 2D FETs have a feature size of 20nm (similar to $L_{TMD}$ of PeFET). Consistent with PeFET based NM/STeP-CiM, $V_{DD} = 0.8V$ and array size of 256×256 is used for SRAM-NM. For PeFET-NM and SRAM-NM, scratchpad memories are accessed row-by-row for performing vector-matrix multiplication (Jain et al., 2020). On the other hand, in STeP-CiM the same is performed by accessing 16 rows of a block simultaneously. We reiterate that the primary distinction between STeP-CiM and PeFET-NM is during compute, while they are identical for memory operations – write and read.

### Area

We present our area analysis of STeP-CiM (or PeFET-NM) and SRAM-NM using thin-cell layout (Khare et al., 2002) based on scalable layout ($F$-based) rules, where $F$ = feature size. In this work, $F$ = 20nm for PeFET and 2D FET based on which SRAMs are designed. We use these rules in conjunction with Intel defined 20nm gate/metal pitch rules (Intel 20nm Lithography). The area of PeFET-NM/STeP-CiM obtained from the layout in Figure 6(C) is $202.5F^2$ while that of SRAM-NM is $378F^2$. We estimate the area of SRAM-NM based on the layout analysis of 2D FET SRAM by (Thakuria et al., 2020). Finally, we report in Figure 11(A) that the layout footprint of PeFET-NM/STeP-CiM is 46% smaller than SRAM-NM.

### Read and Write Comparisons

Performance and energy of STeP-CiM and PeFET-NM are identical since they are essentially the same bit cell during read/write operations, as also discussed earlier. Figure 11(B) indicates that the read latency of STeP-CiM/ PeFET-NM is similar to SRAM-NM. We do not observe faster read in the former despite their compact cell area, since we must account for bit-line charging time in current based sensing mechanism employed during read. In case of SRAM-NM, where we utilize voltage based sensing, this delay may be ignored since $BL/BLB$ are pre-charged to $V_{DD}$.

Next, we elaborate our read energy results. We calculate the read energy in Figure 11(C) considering active energy for 20% utilization, as reported for L2 cache by (Park et al., 2012) and leakage energy for remaining 80% utilization. The active read energy of STeP-CiM/PeFET-NM is 9× higher compared to SRAM-NM. This is because, current based sensing in STeP-CiM/PeFET-NM necessitate switching $BL_1$, $BL_2$ to $V_{DD}/2$ and $RBL_1$, $RBL_2$ to $V_{DD}$ during read, causing energy overheads. In case of SRAM-NM, we utilize voltage based sensing in which $BL/BLB$ discharge by a small voltage of 50mV from their pre-charged state. This incurs low active read energy in SRAM-NM than in current based sensing of STeP-CiM and PeFET-NM. However, leakage energy from the 80% idle utilization dominates in SRAM-NM, while it is insignificant in STeP-CiM/PeFET-NM. This helps reduce the read energy overhead of STeP-CiM/PeFET-NM over SRAM-NM to 55% as shown in Figure 11(C).

Now, we present the write analysis. Due to polarization switching delay in STeP-CiM/PeFET-NM, they show 3.97× higher write time over SRAM-NM (Figure 11(D)).

Interestingly, the write energy of STeP-CiM/PeFET-NM is 18% lower than SRAM-NM (Figure 11(E)). Note that, similar to read, total write energy is reported considering 20% active utilization and 80% leakage in an L2 cache (Park et al., 2012). Although the active energy of STeP-CiM/PeFET-NM is 2× higher than SRAM-NM due to polarization switching, we observe benefits in total write energy due to low utilization rates of modern day caches and dominating leakage energy in SRAM-NM (Park et al., 2012). In this scenario, SRAM-NM is leaking for the remaining 80% utilization, while PeFET-NM/STeP-CiM do not, resulting in overall improvement in the latter. Note that while leakage is a dominant component in SRAMs for memory operation (where a single row is accessed), multiple-row access during MAC operation (discussed next) suppresses this effect.

**MAC**

The highlight of STeP-CiM is that we can access 16 multiple rows parallelly. On the contrary, it needs to be done sequentially in NM baselines. This property benefits both performance and energy of MAC operations using STeP-CiM. Compared to SRAM-NM, we observe ~91% benefits in MAC latency of STeP-CiM, while PeFET-NM shows comparable latency as SRAM-NM (Figure 11(F)).

With respect to MAC energy in Figure 11(G), STeP-CiM shows 15% improvement over SRAM-NM. Note that we obtain benefits in MAC energy with STeP-CiM because of high parallelism mentioned earlier, despite overheads of current sensing. On the contrary, Figure 11(G) shows overhead of write energy of PeFET-NM over SRAM-NM. This is attributed to high energy consumption of current based sensing in the former compared to low energy voltage based sensing. It is important to mention that since > 90% operations in DNNs are MACs, overheads in standard read and write operations are amortized due to significant MAC benefits of the proposed STeP-CiM design. Consequently, large improvements in system performance and energy is observed in STeP-CiM, which we discuss in system level analysis next.

**6.2 System Evaluation**

Here, we evaluate the system-level energy and performance benefits of CiM using STeP-CiM in 5 state-of-the-art DNN benchmarks, *viz.* AlexNet, ResNet34, Inception, LSTM and GRU.

**Simulation Framework**

We design our compute-in-memory (CiM) architecture based on TiM-DNN (Jain et al., 2020) with 32 STeP-CiM arrays, where each array consists of 256×256 STeP-CiM cells, providing a total memory-capacity of 2 Mega ternary words (512 kB). By activating 16 rows simultaneously in each of these arrays, we can perform 8196 parallel vector MAC operations with a vector length of 16. The peripheral circuitry of the STeP-CiM array consists of ADCs (Figure 9) and small compute elements to sense the MAC outputs and perform partial-sum reduction (Jain et al., 2020). We compare the STeP-CiM system with two NM baseline architectures, SRAM-NM and PeFET-NM, constructed with the corresponding memory technologies. We perform the MAC computations and partial-sum reduction in the near-memory compute (NM) units, the inputs to which are read in a sequential row-by-row manner from each memory array. We design two variants of the near-memory baseline – (i) *iso*-capacity and (ii) *iso*-area. The *iso*-capacity SRAM-NM and PeFET-NM baselines contain 32 memory arrays of size 512×256 (identical to STeP-CiM system). We design the *iso*-area baseline architectures with 21 SRAM-NM and 35 PeFET-NM memory arrays, each of size 512×256. We design the SRAM-NM *iso*-area baseline with a smaller number of memory arrays compared to PeFET-NM because SRAM-NM suffers area overhead due to large footprint of SRAM cell. Further, the STeP-CiM array is 1.09× larger in area compared to PeFET-NM due to the area-overhead of the ADCs. We leverage the lower area of PeFET-NM to place a larger number of memory arrays compared to STeP-CiM.

**Performance**

Figure 11(H) shows the performance benefits of STeP-CiM over *iso*-capacity and *iso*-area SRAM-NM and PeFET-NM baselines. We obtain 6.11× and 6.13× average speed-up over the *iso*-capacity

SRAM-NM and PeFET-NM respectively, across the benchmarks considered. Similarly, the average speed-up over *iso*-area SRAM-NM and PeFET-NM is 8.91× and 5.67× respectively. The performance improvements over the near-memory baselines arise from the massively parallel in-memory MAC computation capability of STeP-CiM. The SRAM-NM and PeFET-NM *iso*-capacity baselines have similar performances due to similar memory read latency (discussed in the array level results). Note that, performance enhancement of STeP-CiM over *iso*-area SRAM-NM is greater than over *iso*-capacity SRAM-NM. This is due to higher throughput of STeP-CiM than SRAM-NM at *iso*-area, in addition to the benefits of massively parallel MAC operations. The boosted throughput follows from the larger number of memory arrays of STeP-CiM (32 vs. 21 of SRAM-NM) available for computation at *iso*-area. Contrarily, the performance benefits of STeP-CiM over PeFET-NM at *iso*-area is slightly diminished (relative to the *iso*-capacity case) because PeFET-NM has a comparatively larger number of memory arrays (35 arrays of PeFET-NM compared to 32 of STeP-CiM at *iso*-capacity).

**Energy**

We now present the system-level energy benefits of STeP-CiM compared to near-memory baselines in Figure 11(I). We note that in this evaluation, the *iso*-area and *iso*-capacity baselines are equivalent since the total energy depends on the total number operations that remain the same across these baselines. Therefore, we report the energy benefits of STeP-CiM against the *iso*-area baselines. We achieve 3.2× and 6.07× average energy reduction compared to *iso*-area/capacity SRAM-NM and PeFET-NM respectively for the benchmarks considered. The superior energy efficiency of the proposed STeP-CiM system is due to the parallelism offered by the STeP-CiM arrays as a result of multi-word line assertion for in-memory computation. PeFET-NM consumes higher energy compared to SRAM-NM because of comparatively higher read-energy caused by switching of multiple bit-lines required for current based sensing (as discussed in Section 6.1). We would like to mention here that since the bit-cell for STeP-CiM is reused for PeFET-NM, it is optimized for ternary computation rather than read.

We compare the proposed architecture with existing state-of-the-art ternary DNN accelerators in Table 4. With respect to TeC DNN (Thirumala et al., 2020) and TiM DNN (Jain et al., 2020), we achieve 2.45× and 4.9× improvement in TOPS/W respectively. Moreover, the benefits in TOPS/mm$^2$ are 7× and 15.15× compared to TeC DNN and TiM DNN respectively. The improvements are obtained due to compact size and scaled technology nodes used (20nm vs 45nm and 32nm) and superior compute energy efficiency. Compared to state-of-the-art GPUs, we observe upto 1486× and 5880× in TOPS/W and TOPS/mm$^2$, respectively. Note, however, that the comparisons are made between simulation and experimental results of GPUs.

## 7 Conclusion

In this work we propose a non-volatile memory (STeP-CiM) for ternary DNNs that has the ability to perform signed ternary dot product computation-in-memory. The CiM operation in our design is based on piezoelectric induced dynamic bandgap modulation in PeFETs. We propose a unique technique called Polarization Preserved Piezoelectric Effect Reversal with Dual Voltage Polarity (PiER) which we show is amenable for signed ternary computation-in-memory. Using this property along with multi-word-line assertion, STeP-CiM performs massively parallel dot-product computations between signed ternary inputs and weights. From our array level analysis, we observe 91% lower delay and energy improvement of 15% and 91% for in-memory multiply-and-accumulate operations compared to near-memory approaches designed with 2D FET SRAM and PeFET respectively. Our system-level evaluations show that STeP-CiM achieves upto 6.13× and 8.91× average performance improvement; upto 6.07× and 3.2× reduction in energy compared to PeFET and SRAM based near memory baselines, respectively across 5 state-of-the-art DNN benchmarks.

## 8 Tables

Table 1. Parameters used for in PeFET model

| Parameter | Value | References |
|---|---|---|
| Remnant polarization of PZT-5H, $P_R$ [C/m$^2$] | 0.32 | (Malakooti and Sodano, 2013) |
| Saturation polarization of PZT-5H, $P_S$ [C/m$^2$] | 0.35 | |
| Coercive electric field of PZT-5H, $E_C$ [kV/cm] | 9 | |
| Dielectric constant of PZT-5H, $\epsilon_{r,PE}$ | 4000 | |
| Out-of-plane piezoelectric coupling coefficient of PZT-5H, $d_{33}$ [pm/V] | 650 | |
| In-plane piezoelectric coupling coefficient of PZT-5H, $d_{31}$ [pm/V] | -320 | |
| Polarization switching time, $\tau_{PE}$ [ns] | 1.8 | (Larsen et al., 1991) |
| Thickness of monolayer MoS$_2$, $t_{TMD}$ [nm] | 0.65 | (Peña-Álvarez et al., 2015) |
| Bandgap of monolayer MoS$_2$, $E_0$ [eV] | 1.5 | |
| Coefficient of bandgap change in monolayer MoS$_2$, $\alpha_{TMD}$ [eV/GPa] | 0.800 | |
| Mobility of monolayer MoS$_2$, $\mu_{TMD}$ [cm$^2$/Vs] | 90 | (Hosseini et al., 2015; Yu et al., 2017) |
| Contact resistance, $R_C$ [Ωμm] | 200 | (Schulman et al., 2018) |
| Thickness of PZT-5H, $t_{PE}$ [nm] | 600 | |
| Area of hammer, $A_{PE}$ ($L_{PE} \times W_{PE}$) [nm$^2$] | 100 × 180 | |
| Area of active MoS$_2$/nail beneath MoS$_2$, $A_{TMD}$ ($L_{TMD} \times W_{TMD}$) [nm$^2$] | 30 × 20 | |
| Thickness of nail, $t_{nail}$ [nm] | 10 | |
| Thickness of Al$_2$O$_3$ used as gate oxide, $t_{OX}$ [nm] | 3 | |
| Permittivity of Al$_2$O$_3$ used as gate oxide, $\epsilon_{r,OX}$ [nm] | 12.5 | |
| Length of source/drain contacts, $L_{S/D}$ [nm] | 40 | |
| Supply/drain/write voltage, $V_{DD}$ [V] | 0.8 | |
| Gate voltage during read/compute, $V_{GS}$ [V] | 0.4 | |

Table 2. Summary of bias conditions and PeFET resistance state with *PiERRe* and *PiERCe* modes.

| Mode | Operation | Gate voltage ($V_G = V_R$) | Back contact voltage ($V_B$) | Voltage across PE ($V_{GB}$) | Sensed resistance of PeFET with +P | Sensed resistance of PeFET with -P |
|---|---|---|---|---|---|---|
| *PiERRe* | Read/Compute | 0.4V | 0V | 0.4V = $V_R$ | $E_G \downarrow$, LRS | $E_G \uparrow$, HRS |
| *PiERCe* | Signed Ternary Compute | 0.4V | 0.8V | -0.4V = -$V_R$ | $E_G \uparrow$, HRS | $E_G \downarrow$, LRS |

Table 3. Signed ternary scheme of {-1, 0, 1} in (A) weights (*W*) represented in terms of polarization stored in PeFETs $M_1$ and $M_2$. (B) Sensed states of weights. (C) Inputs (*I*) encoded utilizing biases in word-line (*WL*) and read word-line (*CWL*). Note that the inputs place PeFETs $M_1$ and $M_2$ into different resistance regimes, *PiERCe* and *PiERRe*. (D) Outputs (*O*) used for MAC computation in STeP-CiM. Subtracted currents on read bit-lines $RBL_1$ and $RBL_2$ signify ternary outputs. (E) Truth table of scalar product in signed ternary regime using STeP-CiM.

| (A) Weight (*W*) encoding ||| (B) Read current (*PiERRe* mode) for weights in (A) ||
|---|---|---|---|---|
| $M_1$ | $M_2$ | W | $I_{RBL1}$ (for $M_1$) | $I_{RBL2}$ (for $M_2$) |
| -P | -P | 0 | $I_{HRS}$ | $I_{HRS}$ |
| +P | -P | 1 | $I_{LRS}$ | $I_{HRS}$ |
| -P | +P | -1 | $I_{HRS}$ | $I_{LRS}$ |

| (C) Input (*I*) encoding ||| (D) Output (*O*) encoding in terms of $I_{RBL1}$ - $I_{RBL2}$ ||||
|---|---|---|---|---|---|---|
| WL | CWL | I | $I_{RBL1}$ | $I_{RBL2}$ | $I_{RBL1}$ - $I_{RBL2}$ | O |
| | | | 0 | 0 | 0 | |
| 0 | 0 | 0 | $I_{HRS}$ | $I_{HRS}$ | 0 | 0 |
| | | | $I_{LRS}$ | $I_{LRS}$ | 0 | |
| $V_{DD}$ | 0 | 1 (*PiERRe*) | $I_{LRS}$ | $I_{HRS}$ | **$I_{LRS}$ - $I_{HRS}$** | 1 |
| $V_{DD}$ | $V_{DD}$ | -1 (*PiERCe*) | $I_{HRS}$ | $I_{LRS}$ | **$I_{HRS}$ - $I_{LRS}$** | -1 |

(E) Truth table of scalar product ($I \times W = 0$) in signed ternary regime using STeP-CiM.

| WL | CWL | I | $M_1$ | $M_2$ | W | $I_{RBL1}$ | $I_{RBL2}$ | $I_{RBL1}$ - $I_{RBL2}$ | O |
|---|---|---|---|---|---|---|---|---|---|
| 0 | 0 | 0 | -P | -P | 0 | 0 | 0 | 0 | 0 |
| | | | +P | -P | 1 | | | | |
| | | | -P | +P | -1 | | | | |
| $V_{DD}$ | 0 | 1 (*PiERRe*) | -P | -P | 0 | $I_{HRS}$ | $I_{HRS}$ | 0 | 0 |
| | | | +P | -P | 1 | $I_{LRS}$ | $I_{HRS}$ | $I_{LRS}$ - $I_{HRS}$ | 1 |
| | | | -P | +P | -1 | $I_{HRS}$ | $I_{LRS}$ | $I_{HRS}$ - $I_{LRS}$ | -1 |
| $V_{DD}$ | $V_{DD}$ | -1 (*PiERCe*) | -P | -P | 0 | $I_{LRS}$ | $I_{LRS}$ | 0 | 0 |
| | | | +P | -P | 1 | $I_{HRS}$ | $I_{LRS}$ | $I_{HRS}$ - $I_{LRS}$ | -1 |
| | | | -P | +P | -1 | $I_{LRS}$ | $I_{HRS}$ | $I_{LRS}$ - $I_{HRS}$ | 1 |

Table 4. System level comparison with state of the art DNNs.

|  | **STeP-CiM** | **TeC DNN** | **TiM DNN** | **XORBIN** | **NVIDIA Tesla V100** |
|---|---|---|---|---|---|
| **Reference** | This work | (Thirumala et al., 2020) | (Jain et al., 2020) | (Bahou et al., 2018) | (NVIDIA V100 Tensor Core) |
| **Type of study** | Simulation | Simulation | Simulation | Experimental | Experimental |
| **Technology** | 20nm | 45nm | 32nm | 65nm | 12nm |
| **TOPS/W** | 624 | 255 | 127 | 95 (binary ops) | 0.42 (FP16/32 ops) |
| **TOPS/mm$^2$** | 882 | 122 | 58.2 | 3.5 | 0.15 |

## 9 Figures

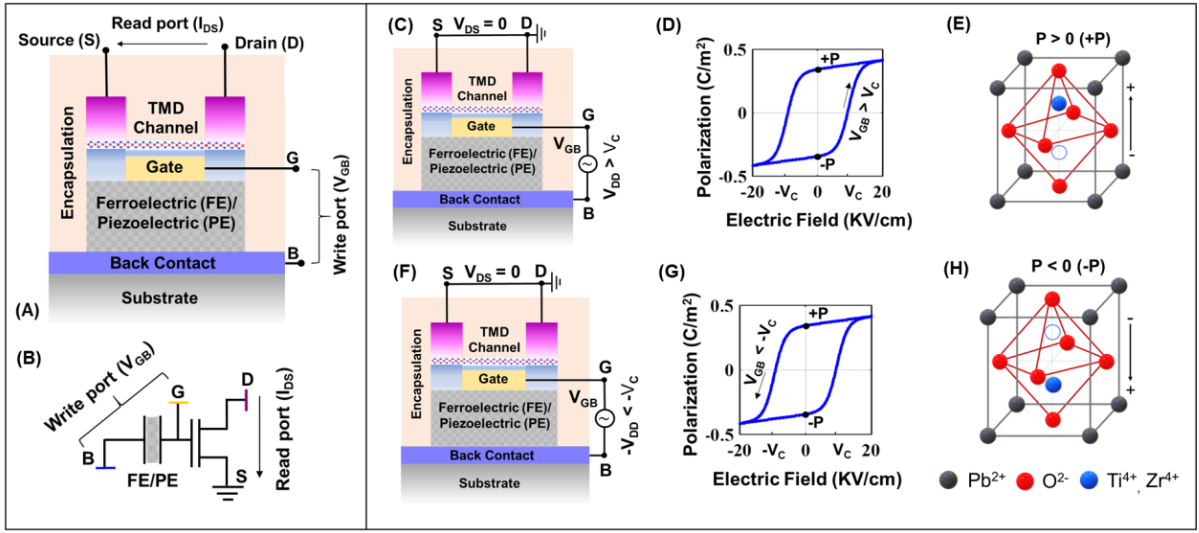

Figure 1. (A) Device structure of PeFET showing its four terminals, namely back (*B*), drain (*D*), gate (*G*) and source (*S*). Ferroelectric or piezoelectric (FE/PE) material that extends non-volatile feature to PeFET is placed between *G* and *B* terminals. (B) Schematic of PeFET. Ferroelectric based storage in PeFET (C) Bias conditions of PeFET for +*P* polarization switching. (D) Polarization vs. electric field response showing +*P* switching behavior for $V_{GB} > V_C$ in PZT-5H, extracted from experimental result by (Malakooti and Sodano, 2013). (E) Structure of PZT in +*P* stable state. (F) Bias conditions of PeFET for -*P* polarization switching. (G) Polarization vs. electric field response showing -*P* switching behavior for $V_{GB} < -V_C$ in PZT-5H, extracted from experimental result by (Malakooti and Sodano, 2013). (H) Structure of PZT in -*P* stable state.

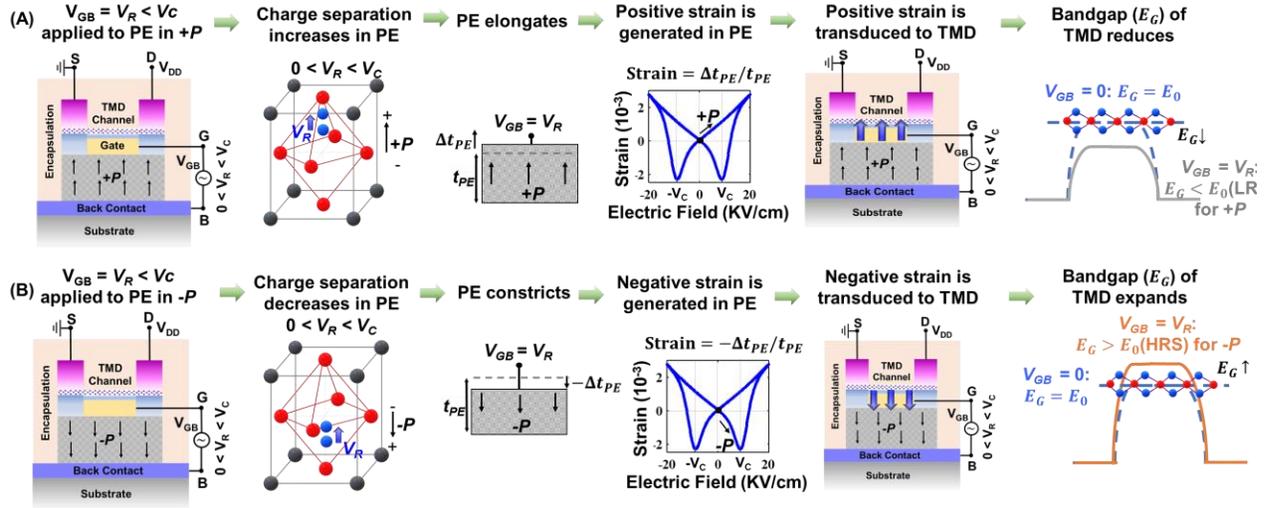

Figure 2. Mechanism of polarization-dependent strain transduction in PE that enables read operation in PeFET. (A) PeFET with +$P$ showing bandgap reduction and (B) PeFET with -$P$ showing bandgap expansion from intrinsic position on application of read voltage that is $0 < V_R < V_C$. The baseline for bandgap comparison is PeFET with $V_{GB} = 0$.

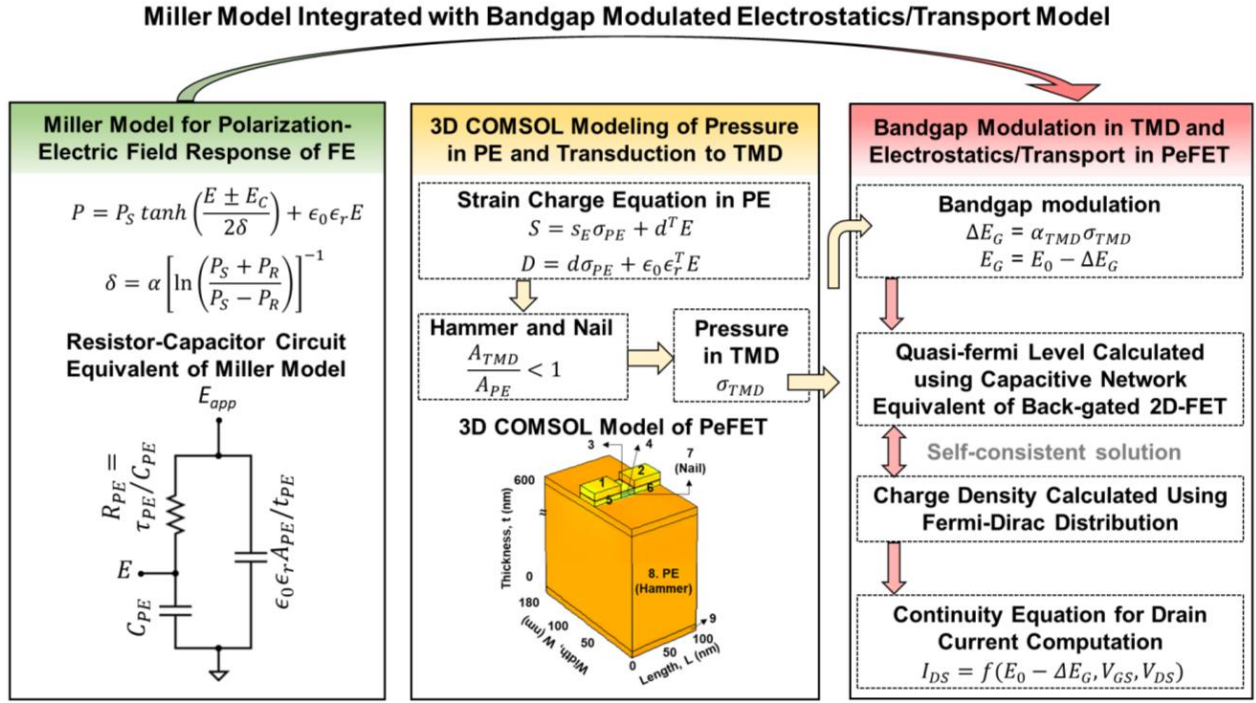

Figure 3. HSPICE compatible model of PeFET formed by integrating (i) Miller model of polarization-electrical field behavior of PE, (ii) 3D COMSOL model of PeFET simulating pressure in PE and its transduction to TMD on application of gate voltage and (iii) electrostatics and transport model of TMD augmented with bandgap modulation behavior.

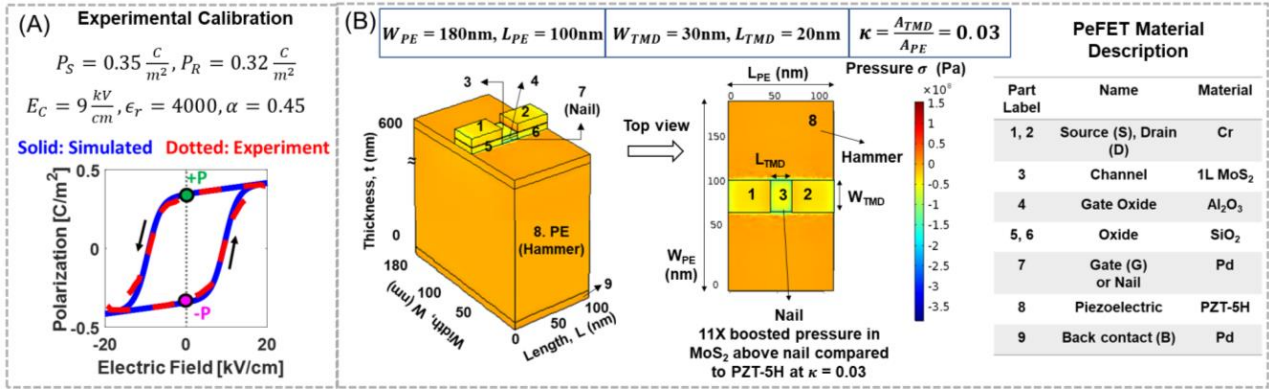

Figure 4. (A) Calibration of simulated polarization-electrical response with experiments. (B) 3D COMSOL model of PeFET with components labelled, their material specification and top view of PeFET signifying hammer which is the top surface area of PE (8) and nail which is the gate below active $MoS_2$ (7). Pressure in $MoS_2$ in the active/nail region is ~11× higher than PE.

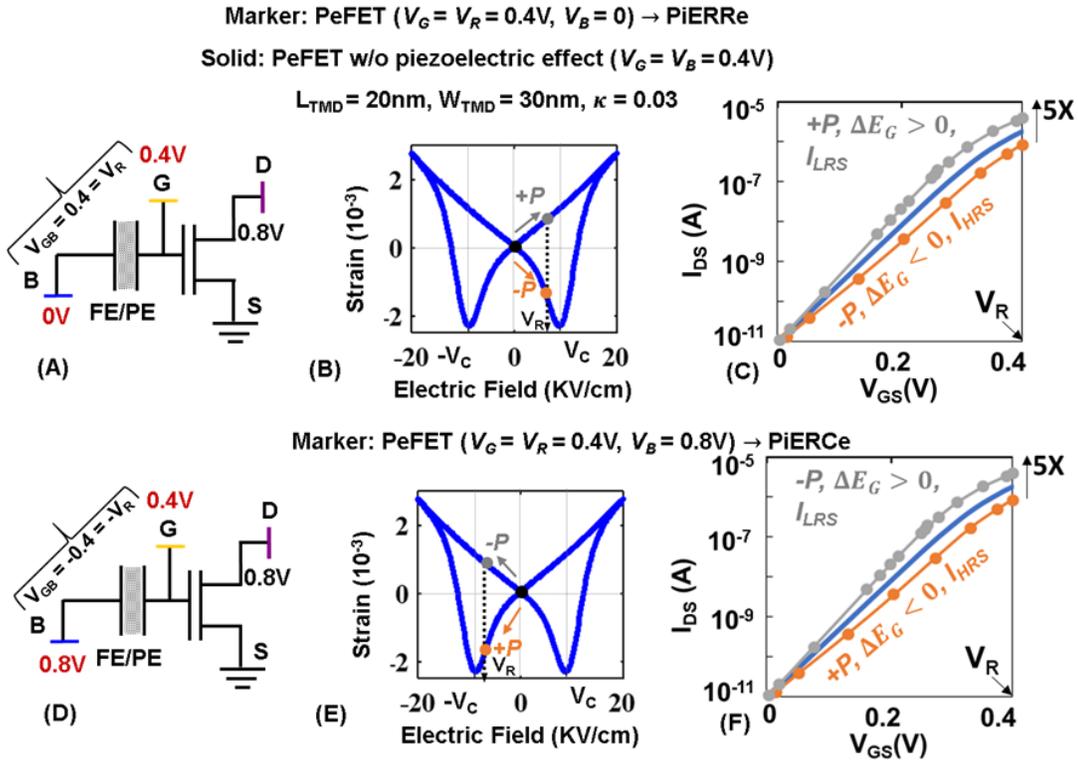

Figure 5 (A) Schematic of PeFET for $V_G$ = 0.4V, $V_B$ = 0V, $V_{GB}$ = 0.4V > 0 or *PiERRe* mode. (B) Strain-Electric field response for $V_{GB} = V_R > 0$. (C) Device characteristics for $V_{GB} > 0$. (D) Schematic of PeFET for $V_G$ = 0.4V, $V_B$ = 0.8V, $V_{GB}$ = -0.4V < 0 or *PiERCe* mode. (E) Strain-Electric field response for $V_{GB} = -V_R < 0$. (F) Device characteristics for $V_{GB} < 0$.

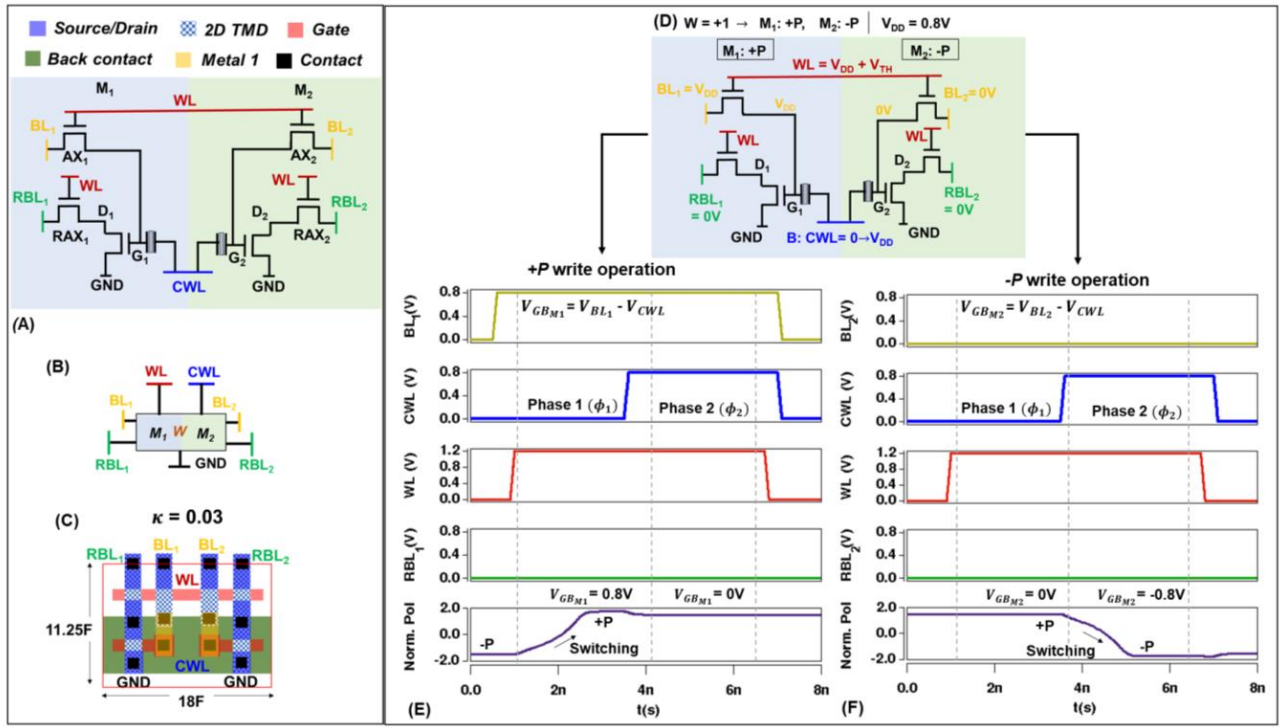

Figure 6 Non-volatile memory cell for STeP-CiM showing (A) Schematic. (B) Symbol. (C) Layout. (D) Schematic of STeP-CiM cell with an example of biases for write (ternary 1). Two-phase write operation depicting in (E) $-P \rightarrow +P$ polarization switching when $V_{GB} = 0.8V > V_C$ and (F) $+P \rightarrow -P$ polarization switching when $V_{GB} = -0.8V < -V_C$. $+P$ and $-P$ states being written to $M_1$ and $M_2$ constitute ternary storage of $W = +1$.

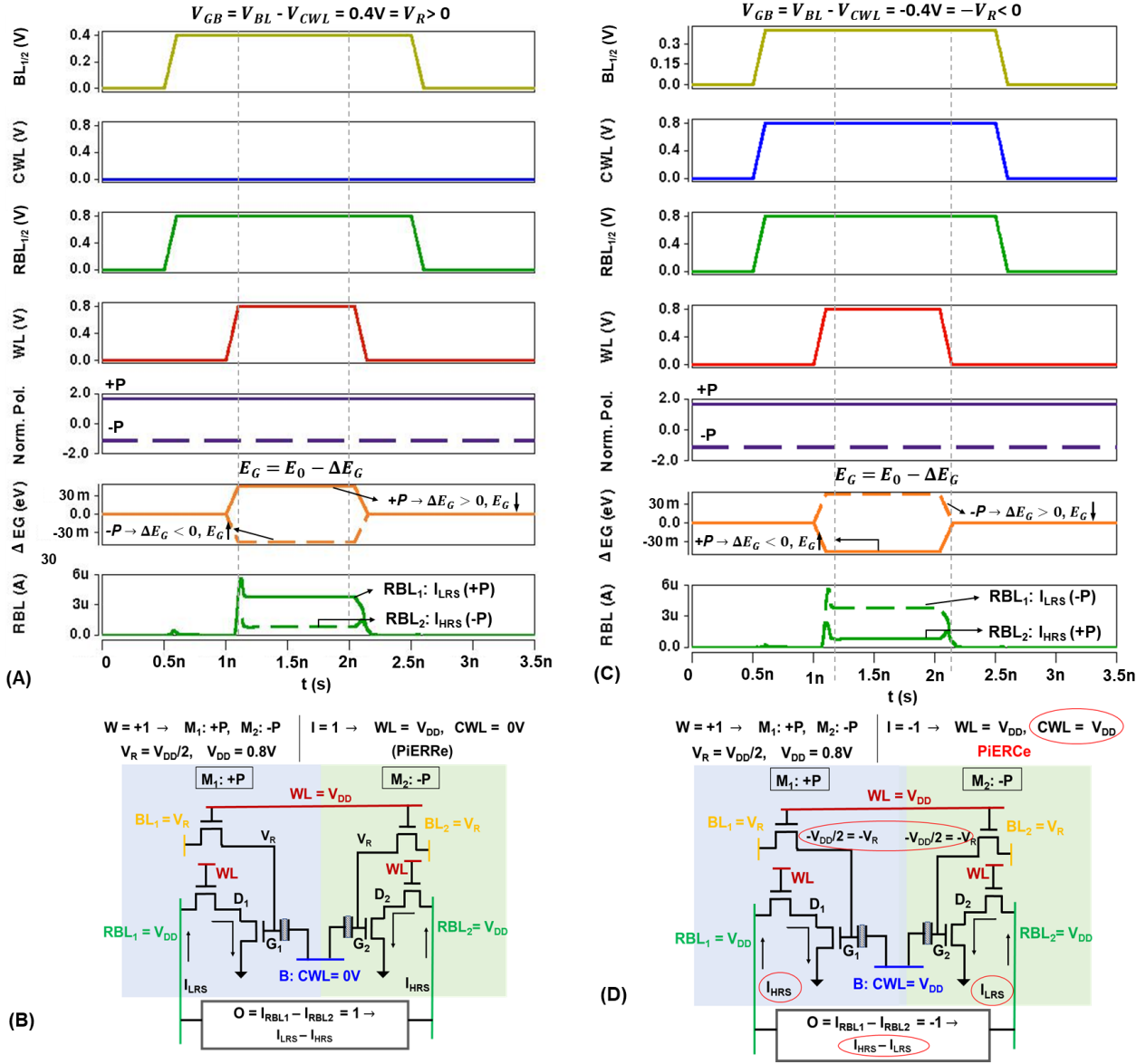

Figure 7. Waveform of (A) read operation showing $I_{LRS}$ for +P and $I_{HRS}$ for -P when $V_{GB} = V_R$ (*PiERRe* mode). This waveform and schematic in (B) show scalar product with $W = 1$ (+P in $M_1$ and -P in $M_2$) and $I = 1$ (corresponding to $CWL = 0V$ and $WL = V_{DD}$), resulting in $I_{LRS}$ on $RBL_1$ due to +P in $M_1$ and $I_{HRS}$ on $RBL_2$ due to -P in $M_2$. (C) Waveform and schematic in (D) depicting scalar product with $W = 1$ (+P in $M_1$ and -P in $M_2$) and $I = -1$ (in *PiERCe* mode corresponding to $CWL = V_{DD}$ and $WL = V_{DD}$). This results in $I_{HRS}$ on $RBL_1$ due to +P in $M_1$ being sensed as HRS due to $I = -1$ and $I_{LRS}$ on $RBL_2$ due to -P in $M_2$ being sensed as LRS due to $I = -1$. For the same weights, currents on $RBL_1$ and $RBL_2$ for $I = 1$ (waveform A) are swapped for $I = -1$ (waveform B) due to difference in *CWL* (highlighted with red circles).

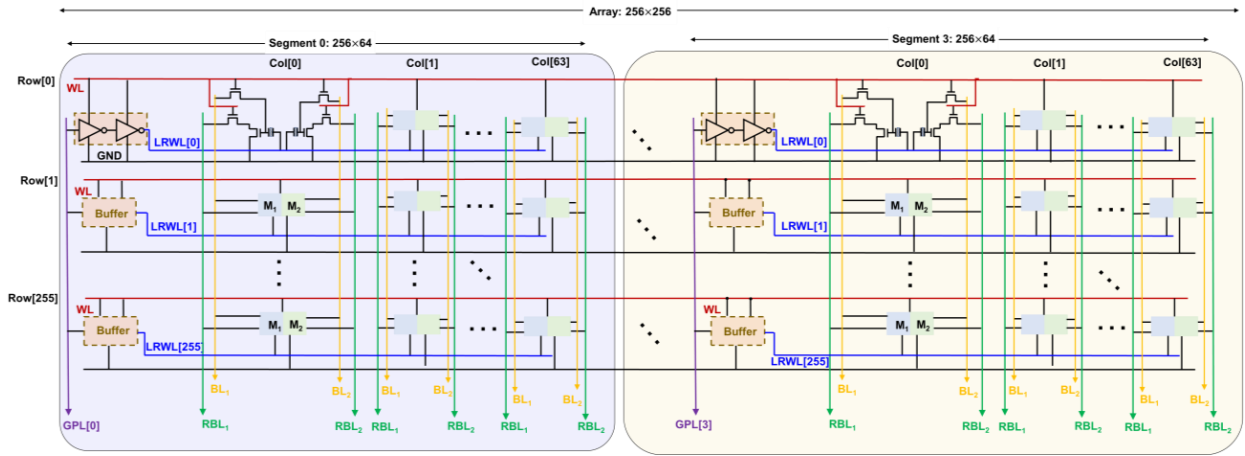

Figure 8 Segmented array of STeP-CiM of size 256×256, comprising of 4 segments, each of size 64×256.

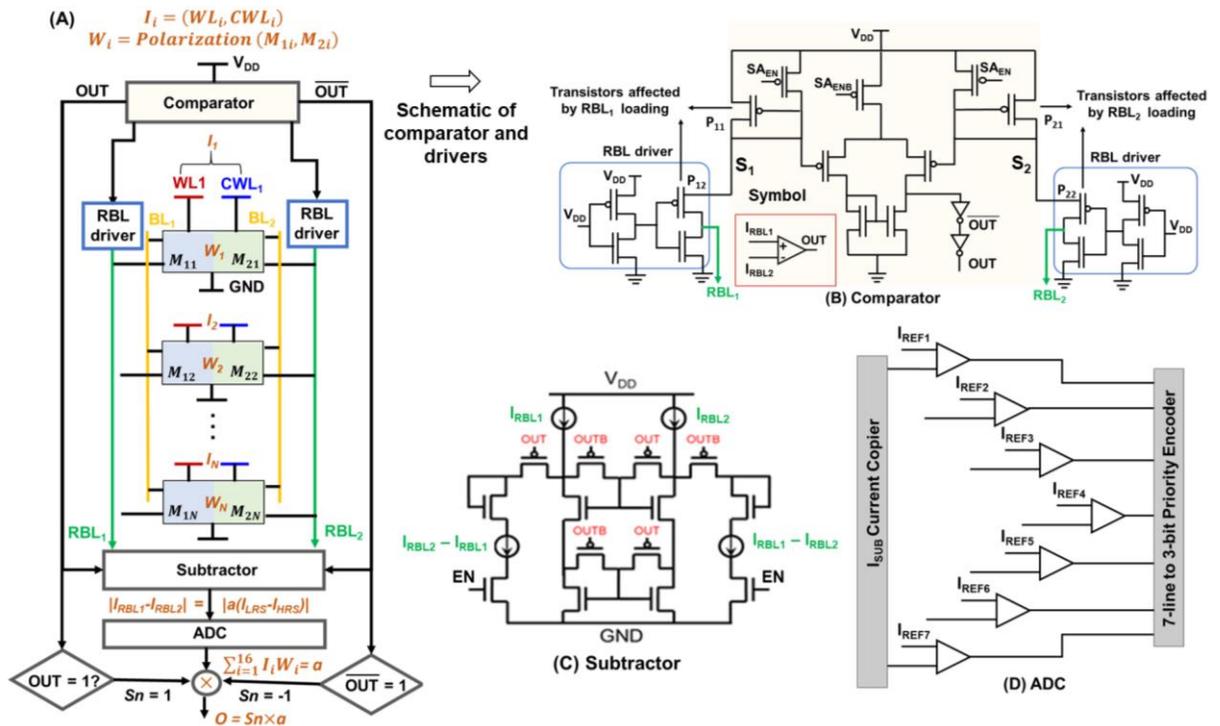

Figure 9 (A) Column of STeP-CiM cell for MAC operation. In the schematic, we label weights (*W*) as combination of polarization in $M_{1i}$ and $M_{2i}$ conforming to encoding in Table 3(A), while inputs of a cell are encoded utilizing *WL* and *CWL* according to Table 3(C). Peripherals used in the simulation of a MAC operation such as comparator, read driver, subtractor and ADC are shown. Schematic of (B) Read drivers interacting *RBL* and comparator. (C) Subtractor and (D) 3-bit ADC represented using symbol in inset of (B).

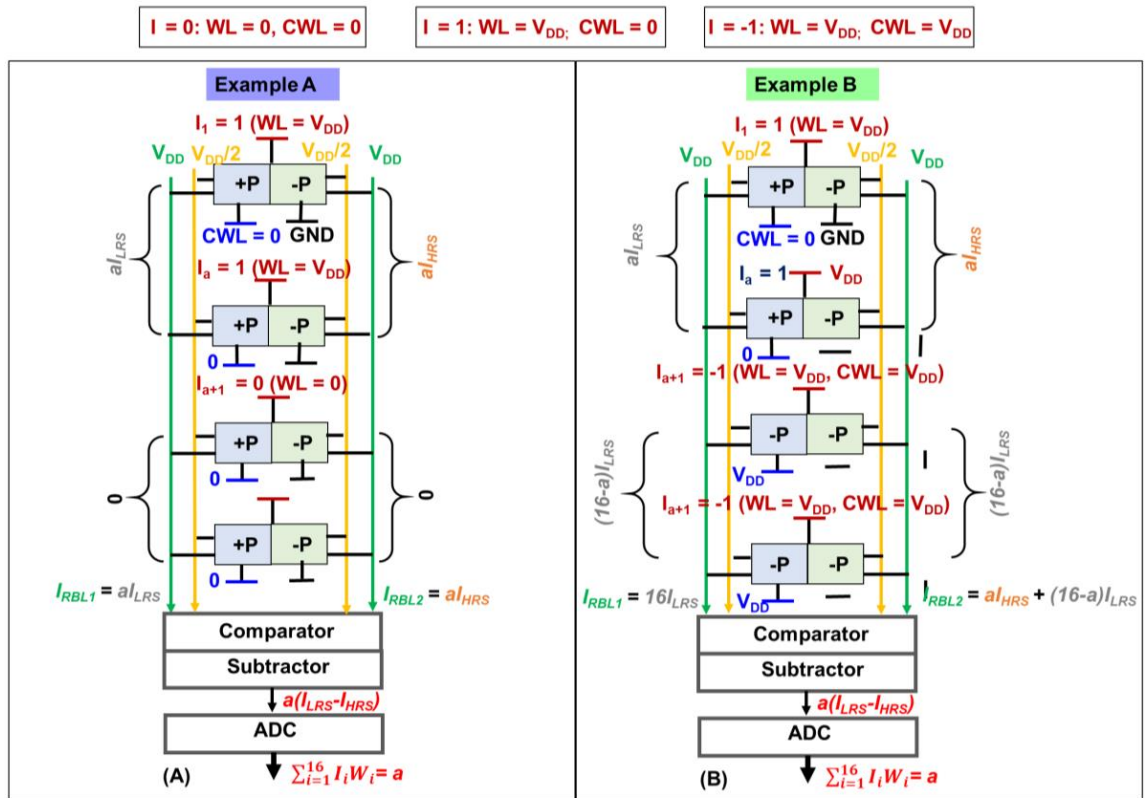

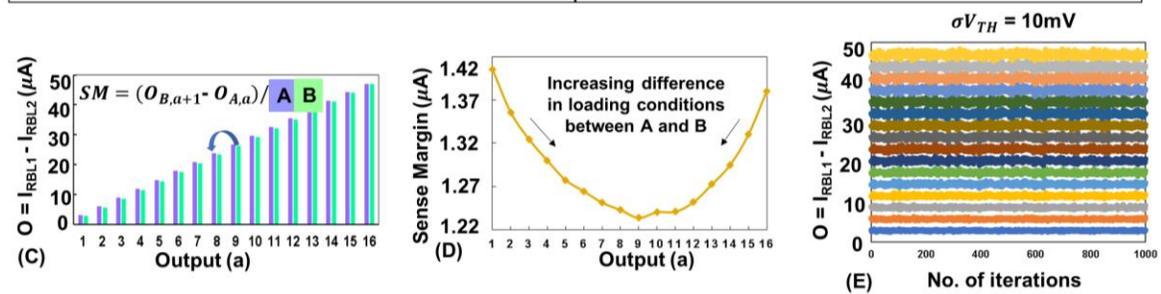

Figure 10 Examples of STeP-CiM for (A) best case (B) worst case loading effect (C) computation of sense margin (D) sense margin (E) variation analysis for 16 activations.

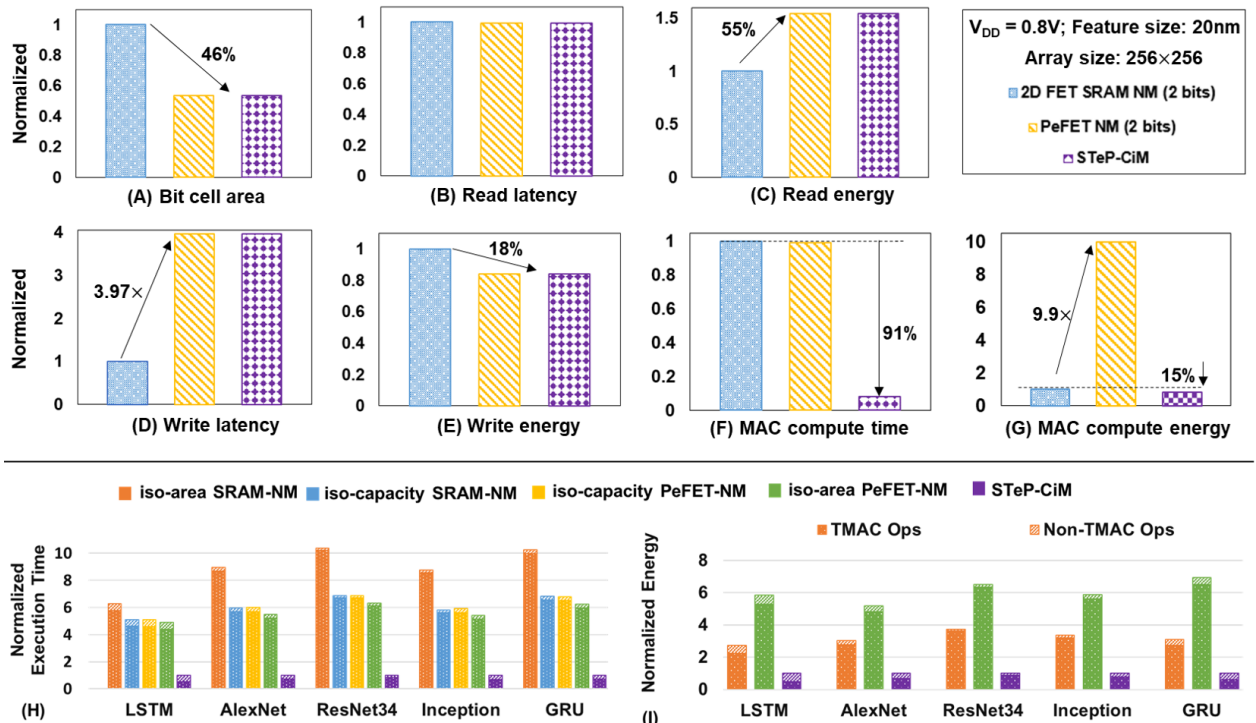

Figure 11 Array level results of STeP-CiM vs. PeFET-NM and SRAM-NM shown for (A) Bit cell area. (B) Read latency. (C) Read energy. (D) Write latency. (E) Write energy. (F) MAC compute time. (G) MAC compute energy. System level results for (H) normalized execution time. (I) Normalized energy of STeP-CiM with respect to *iso*-capacity and *iso*-area PeFET-NM and SRAM-NM.

**Conflict of Interest**

*The authors declare that the research was conducted in the absence of any commercial or financial relationships that could be construed as a potential conflict of interest.*

**Acknowledgements**

This research was supported, in part, by Army Research Office (W911NF-19-1-048) and SRC/NSF funded E2CDA program (1640020).

**Author Contributions**

N.T. and S.K.G. conceived the idea and designed the analysis. S.K.T. contributed to the idea of current based sensing. N.T. and R.E. performed the device, array and system level simulations and analyses. N.T., R.E. and S.K.G. wrote the manuscript. N.T., R.E., S.K.T., A.R. and S.K.G. analyzed the data, discussed the results, agreed on their implications, and contributed to the preparation of the manuscript. A.R. and S.K.G. supervised the project.